\documentclass{JHEP}

\usepackage{amssymb}

\usepackage{amsfonts}

\usepackage{amsbsy}

\usepackage{epsfig}

\newcommand{\im}{\mathop{\rm Im}\nolimits}

\newcommand{\NP}[1]{Nucl.\ Phys.\ {\bf #1}}

\renewcommand{\cal}{\mathcal}

\def\IC{\mathbb{C}}
\def\IZ{\mathbb{Z}}
\def\IR{\mathbb{R}}

\def\IQ{\mathbb{Q}}
\def\CP{\mathbb{CP}}
\def\one{{\bf 1}}
\def\sN{{\bf N}}

\def\sx{{\bf x}}

\def\sr{{\bf r}}

\def\sC{{\bf C}}
\def\sA{{\bf A}}
\def\sJ{{\bf J}}

\def\CD{{\cal D}}

\def\CN{{\cal N}}
\def\CM{{\cal M}}

\def\tr{\mbox{Tr}}
\def\ssigma{{\boldsymbol \sigma}}

\def\s*{\boldsymbol{*}}

\def\CO{{\cal O}}

\newcommand{\be}{\begin{equation}}
\newcommand{\ee}{\end{equation}}
\newcommand{\bea}{\begin{eqnarray}}
\newcommand{\eea}{\end{eqnarray}}

\def\CM{\cal M}

\title{Quantum Quivers and Hall/Hole Halos}

\author{Frederik Denef \\ Department of Mathematics,
Columbia University \\ New York, NY 10027, USA\\
\email{denef@math.columbia.edu}}

\preprint{{\tt hep-th/0206072}}

\abstract{Two pictures of BPS bound states in Calabi-Yau
compactifications of type II string theory exist, one as a set of
particles at equilibrium separations from each other, the other as
a fusion of D-branes at a single point of space. We show how
quiver quantum mechanics smoothly interpolates between the two,
and use this, together with recent mathematical results on the
cohomology of quiver varieties, to solve some nontrivial ground
state counting problems in multi-particle quantum mechanics,
including one arising in the setup of the spherical quantum Hall
effect, and to count ground state degeneracies of certain dyons in
supersymmetric Yang-Mills theories. A crucial ingredient is a
non-renormalization theorem in $\CN=4$ quantum mechanics for the
first order part of the Lagrangian in an expansion in powers of
velocity.}

\begin{document}

%%%%%%%%%%%%%%%%%%%%%%%%%%%%%%%%%%%%%%%%%%%%%%%%%%%%%%%%%%%%%%%%%%%%%%%%
\section{Introduction}\label{sec:intro}
\setcounter{equation}{0}
%%%%%%%%%%%%%%%%%%%%%%%%%%%%%%%%%%%%%%%%%%%%%%%%%%%%%%%%%%%%%%%%%%%%%%%%

One of the big successes of string theory has been the counting of
black hole microstates using D-brane constructions, starting with
\cite{SV}. The basic idea is that black holes, i.e.\ solutions of
the low energy supergravity theory, have a dual description as
D-branes, which provide a reliable description of the physics in
the limit of vanishing string coupling constant, $g_s \to 0$.
Under the assumption that the number of states does not change
when $g_s$ is sent to zero, one can thus count black hole states
by counting D-brane states, and one finds among other things a
precise match with the Bekenstein-Hawking entropy formula. One way
to obtain conservation of the number of states under change of
$g_s$ is to consider BPS black holes and D-brane states, since the
number of supersymmetric ground states of a system is often
invariant under continuous changes of the coupling constants, or
at least well under control.

In this paper we will push further this idea to the counting of
ground state degeneracies of objects more complicated than black
holes. In four dimensional $\CN=2$ supergravities and gauge
theories, there exists BPS states which are multicentered
composites, bound together like atoms in a molecule
\cite{branessugra,DGR,LY,RSVV,argyres}. These $\CN=2$ theories
arise or can be geometrically engineered \cite{geomeng} as certain
Calabi-Yau compactifications of type II string theory, in which
BPS states appear in the limit of vanishing string coupling
constant as wrapped D-branes, localized at a single position in
space. Again there is a duality between these D-branes and the
multi-centered solutions of the effective low energy theory, and
one can follow the same reasoning as for ordinary black holes to
count quantum ground state degeneracies.

The constituents of these multicentered composites do not
necessarily have to be black holes to get interesting counting
problems; they can be ordinary particles. The quantum dynamics of
these systems is therefore often better under control, in both the
multi-particle and the single D-brane pictures, making it possible
to count states on both sides and compare the two. Using the
quiver description of the wrapped D-branes
\cite{DMquiv,DFR,DFR2,FM,DD,Dcat,fiol} (in regimes where this
description is valid), it becomes even possible to follow in
detail the transition from one picture to the other.

One of these interesting counting problems which arises naturally
in this framework is to find the lowest Landau level degeneracies
of a ``quantum Hall halo''. A Hall halo can be thought of as a
charge $\kappa$ magnetic monopole surrounded by a number $N$ of
electrons bound to a sphere of fixed radius around the monopole.
This system had been studied in detail in the condensed matter
literature \cite{qhall}. The generating function $G(t)=\sum_L n_L
\, t^L$ for the number of ground states $n_L$ with spin $J_3 = L/2
- N(\kappa-N)/2$ was found to be
\begin{equation} \label{genfunctintro}
 G(t) = \frac{\prod_{j=1}^\kappa (1-t^{2j})}
 {\prod_{j=1}^N (1-t^{2j}) \prod_{j=1}^{\kappa-N} (1-t^{2j})} \, .
\end{equation}
Through the correspondence with microscopic D-brane states studied
in this paper, this condensed matter problem gets mapped to
counting supersymmetric ground states of quantum mechanics on the
moduli space of a quiver with two nodes, $\kappa$ arrows and
dimension vector $(1,N)$ (see section \ref{quivmath} for
definitions). This moduli space is the Grassmannian
$Gr(N,\kappa)$, i.e.\ the space of $N$-dimensional planes in
$\IC^{\kappa}$. The supersymmetric ground states are in one-to-one
correspondence with the cohomology of this space, which is
classically known. The generating function for the betti numbers
$b_L$, also known as the Poincar\'e polynomial, is \cite{bott}:
\begin{equation}
  P(t) \equiv \sum_L b_L \, t^L = \frac{\prod_{j=1}^\kappa (1-t^{2j})}
 {\prod_{j=1}^N (1-t^{2j}) \prod_{j=1}^{\kappa-N} (1-t^{2j})} \, .
\end{equation}
The cohomology is organized in Lefschetz SU(2) multiplets, which
as we will see coincide here with the spatial spin multiplets,
implying that an $L$-form has spin $J_3 = L/2 - N(\kappa-N)/2$.
Comparing this to (\ref{genfunctintro}), we see that we have
indeed exact agreement.

The main part of this paper is aimed at obtaining an understanding
of why these two counting problems, and many generalizations
thereof, are equivalent. We achieve this by modeling the systems
under consideration as $\CN=4$ supersymmetric quiver quantum
mechanical systems, obtained as the dimensional reduction of four
dimensional $\CN=1$ gauge theories. These quantum mechanical
models have both ``Higgs'' and ``Coulomb'' branches, where the
Higgs branch is the one supporting the microscopic single D-brane
picture of bound states, while the Coulomb branch supports the
multi-centered picture. Classically, the Coulomb branch is
trivially flat, but we will see that quantum effects induce an
effective potential and magnetic interaction, which are precisely
of the form needed to get multicentered ``molecular'' BPS bound
states. The match in this respect between the supergravity and
substringy regimes will be traced back to a non-renormalization
theorem for the term linear in ``velocities'' in the $\CN=4$
multiparticle Lagrangian. We will furthermore identify a regime in
which the state lives essentially on the Higgs branch, with
quantum fluctuation effectively washing out the structure on the
Coulomb branch, and a complementary regime in which the opposite
happens. Lowering $g_s$ down to zero corresponds then to
``squeezing'' the states from their life on the Coulomb branch to
a new life on the Higgs branch.

The counting problem on the Higgs branch reduces to finding the
Betti numbers of the cohomology of the associated quiver moduli
space. For quivers without closed loops, this problem was recently
solved in full generality \cite{reineke}, allowing us to do
predictions of quantum ground state degeneracies of various highly
nontrivial systems of generalized interacting
``electron-monopole'' systems, for which even the classical
configuration moduli space can be extremely complicated. This
includes ground state degeneracies of certain dyons in
supersymmetric Yang-Mills theories.

Finally, this work also provides new insight in the correspondence
between stability of multi-centered BPS configurations
\cite{branessugra,DGR} and stability of wrapped D-branes
\cite{DFR}-\cite{fiol},\cite{joyce}-\cite{DGJT}.

The organization of this paper is as follows. In section
\ref{sec:particles}, we analyze in detail the structure of the
molecular multi-centered bound states, both in supergravity and in
abstract generality, and (re-)establish a supersymmetric
non-renormalization theorem which fixes the part of the Lagrangian
responsible for this kind of bound states. In section
\ref{sec:micro}, we introduce the quiver model, recall some of its
mathematical features, and give its physical interpretation. In
section \ref{sec:qq}, we turn to the quantization of this model
and explain how the two pictures of bound states are related in
this framework. The case of the Hall halo, as summarized above, is
studied in more detail as a non-trivial example of the
correspondence. In section \ref{sec:LLL} we present more tests and
applications. In particular, we formulate some predictions of
ground state degeneracies of generalizations of the Hall halo, and
reproduce and refine the ground state counting of the Stern-Yi
dyon chain \cite{sy}. Our conclusions and open questions are
discussed in section \ref{sec:conclusions}. The appendices give
our conventions and various explicit expressions for quiver
Lagrangians and their supersymmetries.

%%%%%%%%%%%%%%%%%%%%%%%%%%%%%%%%%%%%%%%%%%%%%%%%%%%%%%%%%%%%%%%%%%%%%%%%
\section{BPS bound states of particles in 3+1 dimensions}\label{sec:particles}
\setcounter{equation}{0}
%%%%%%%%%%%%%%%%%%%%%%%%%%%%%%%%%%%%%%%%%%%%%%%%%%%%%%%%%%%%%%%%%%%%%%%%

Consider a 3+1 dimensional $\CN=2$ supergravity theory containing
a number of massless abelian vector multiplets coupled to a number
of BPS particles with arbitrary corresponding electric and
magnetic charges. Systems like this typically arise in the low
energy limit of Calabi-Yau compactifications of type II string
theory, where D-branes wrapped around nontrivial cycles manifest
themselves as charged particles in the 3+1 dimensional low energy
effective theory.

These particles have long distance interactions through their
coupling to the metric, the vector fields, and the complex scalars
of the vector multiplets. The relative strength of these forces
depends on the choice of vacuum (expectation values of the complex
scalars). The force between static BPS particles with proportional
charge vectors always vanishes, but, for generic vacua, this is
not true for particles with non-proportional charges. Moreover,
the interactions are sufficiently complex to allow situations
where nontrivial balancing between the different forces occurs at
certain separations. Many of the resulting classical bound states
are BPS, as studied in detail in \cite{branessugra,DGR}. In
quantum field theories without gravity, similar structures emerge,
see for instance \cite{LY,RSVV,argyres}.

In this section, we will investigate such interacting
multiparticle systems. We will take two approaches. The first one
is based on solutions of the supergravity theory, and the second
one on the constraints imposed by supersymmetry on the particle
mechanics itself. The detailed supergravity picture is not really
essential for the main purpose of this paper, but as it gives a
concrete physical realization of these systems (and was in fact
the original inspiration for this work), we include it here,
though we will only briefly review the main features, referring to
\cite{branessugra,DGR} for details.

Let us first recall a few basic notions. Assume the supergravity
theory under consideration has $n$ $U(1)$ vectors apart from the
graviphoton. A charge vector $Q$ is then specified by $2n+2$
integers: $n+1$ electric charges $Q_{e,I}$ and $n+1$ magnetic
charges $Q_m^I$. Two charges $Q$ and $\tilde{Q}$ are called
mutually nonlocal if their Dirac-Schwinger-Zwanziger product
$\langle Q,\tilde{Q} \rangle \equiv Q_{m,I} \tilde{Q}_e^I - Q_e^I
\tilde{Q}_{m,I}$ is nonzero. In the IIB wrapped D3-brane picture,
the geometric interpretation of this product is the usual
intersection product: it counts (with signs) the number of
intersection points of the corresponding two 3-branes. A pair of
mutually nonlocal charges can be thought of as a (generalized)
monopole-electron system.\footnote{We will use the term
``monopole-electron system'' loosely in this paper. A more precise
phrasing would be ``a pair of BPS particles which have magnetic
resp.\ electric charge with respect to the same $U(1)$, upon a
suitable choice of charge basis''. \label{fn:monelectr}}

A central role in the description of BPS states is played by the
central charge. This is a function of the complex vector multiplet
scalars $z^a$, $a=1,\ldots,n$ (which are the complex structure or
the complexified K\"ahler moduli of the CY in the IIB resp.\ IIA
string theory context), and a linear function of the
electromagnetic charge $Q$. We will denote it by $Z_Q(z)$. The
dependence on $z$ is holomorphic up to an overall normalization
factor (the exponential of half the K\"ahler potential on the
vector moduli space). The mass of a BPS particle of charge $Q$, in
a vacuum specified by the vevs $\langle z^a \rangle =
z^a|_{r=\infty} \equiv u^a$ is given by $M=|Z_Q(u)|/l_P$, where
$l_P$ is the four dimensional Planck length, defined here as the
square root of the Newton constant. The interpretation of the
phase of the central charge is the embedding angle of the residual
$\CN=1$ supersymmetry in the original $\CN=2$ \cite{M}.

\subsection{Single probe in supergravity background}
\label{sec:sugraprobe}

We start with the simplest case: a (light) probe BPS particle of
($n+1$-component) charge $q$ in the background produced by another
(heavy) BPS particle of charge $Q$, fixed at the origin. We assume
the particles are mutually nonlocal, i.e.\ $\langle q,Q \rangle
\neq 0$. With $u^a=z^a|_{r=\infty}$, the mass of the probe is
$m=|Z_q(u)|/l_P$.

The metric produced by the fixed source is of the form
$ds^2=-\rho^2 dt^2 + \rho^{-2} d{\sx}^2$. The redshift factor
$\rho$ as well as the vector multiplet scalars $z^a$ are function
of the coordinate distance $r=|\sx|$ only, and obtained as the
solutions to the integrated BPS equations of motion (first derived
in \cite{FKS} and written in the following form in
\cite{branessugra}):
\begin{equation} \label{floweq}
 \left. 2 \, \rho^{-1} \im[e^{-i\alpha} Z_{Q'}(z)]\right|_r = - \frac{l_P \, \langle Q',Q
\rangle}{r}
 + \left. 2 \im[e^{-i \alpha} Z_{Q'}(z)]\right|_{r=\infty}
\end{equation}
for arbitrary charges $Q'$ (or equivalently for a basis of
$2(n+1)$ charges $Q'$), with $\alpha \equiv \arg Z_Q$. For most
Calabi-Yau manifolds, because of the complicated dependence of the
central charges $Z$ on the moduli $z^a$, it is in general not
possible to find exact analytic solutions for $z^a(r)$ and
$\rho(r)$, but several approximate analytical and numerical
solutions have been obtained, and many properties can be infered
directly from the equations. As noted before, the details of this
are not important for the purpose of this paper.

The action for a BPS probe is \cite{BBS,D0}
\begin{equation}
 S = - l_P^{-1} \int |Z_q| \, ds + \int \langle q, {\cal A} \rangle \, ,
\end{equation}
where ${\cal A}$ is the ($n+1$-component) electromagnetic
connection of the background. Using the BPS equations of motion
(\ref{floweq}) for $z^a(r)$ and $\rho(r)$ and those for ${\cal A}$
(for which we refer to \cite{branessugra}), one gets a probe
action of the form $S = \int (K - V + M) \, dt$ with kinetic,
potential and magnetic parts given by \cite{branessugra}:
\begin{eqnarray}
 K &=& - l_P^{-1} \, \rho \, |Z_q| \, \left( \sqrt{1 - \rho^{-4} \, \dot{\sx}^2} - 1 \right) \label{Kformula} \\
  &\approx& l_P^{-1} \, \rho^{-3} |Z_q| \, \dot{\sx}^2 / 2 \quad
   \mbox{if } \rho^{-2} |\dot{\sx}| \ll 1 \, , \label{KformulaNR} \\
 V &=& l_P^{-1} \, \rho \, |Z_q| \, (1-\cos(\alpha_q-\alpha)) \\
   &=& 2 \, l_P^{-1} \, \rho \, |Z_q| \, \sin^2 [(\alpha_q-\alpha) /2] \label{Vformula}\\
   &\approx& l_P^{-1} \, \rho \, |Z_q| (\alpha_q-\alpha)^2 \quad \mbox{if }
   |\alpha_q -\alpha| \ll 1 \, , \label{VformulaNR} \\
 M &=& \frac{1}{2} \, \langle q,Q\rangle \, \sA^d \cdot \dot{\sx}
 \, .\label{Mformula}
\end{eqnarray}
The dot denotes $d/dt$, $\alpha_q$ and $\alpha$ are the phases of
$Z_q$ resp.\ $Z_Q$, and $\sA^d$ is a $U(1)$ vector potential for
the Dirac magnetic monopole carrying one flux quantum:
\begin{equation} \label{dirpot}
 \sA^d \cdot \dot{\sx} = \frac{1}{2} (\pm 1 - \cos \vartheta) \, \dot{\varphi}
 = \frac{1}{2} \, (\pm 1 - z/r) \, \frac{x \dot{y} - y \dot{x}}{x^2+y^2} \, .
\end{equation}
The approximations (\ref{KformulaNR}) and (\ref{VformulaNR})
correspond to the nonrelativistic limit (i.e.\ kinetic and
interaction energies much smaller than the total mass of the
system). Note that the various quantities appearing in
(\ref{Kformula}) - (\ref{Mformula}) are $r$-dependent through the
$r$-dependence of $\rho$ and $z$. Fig.\ \ref{potential} shows some
typical potentials $V(r)$; concrete examples for compactifications
on the Quintic can be found in \cite{branessugra,DGR}.

\FIGURE[t]{\centerline{\epsfig{file=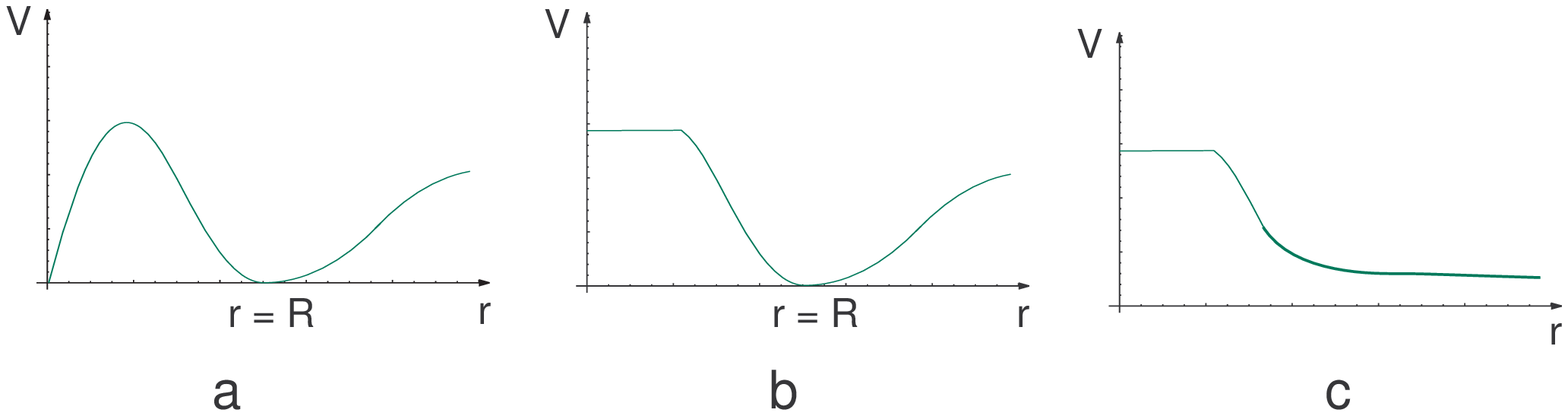,height=4.5cm}}
\caption{Typical examples of a potential for a test particle in
the background field of another charge. The potential is taken to
be zero when the total energy saturates the BPS bound. (a)
corresponds to a black hole source (with horizon at $r=0$), (b)
and (c) to a source with an enhan\c{c}on-like \cite{enhancon} core
instead of a horizon (an ``empty hole'' in the terminology of
\cite{branessugra}). The cases (a) and (b) have $\langle q,Q
\rangle \sin(\alpha_q-\alpha)_{r=\infty}>0$, while (c) has
$\langle q,Q \rangle \sin(\alpha_q-\alpha)_{r=\infty}<0$.
}\label{potential}}

If at a certain radius $r=R$ the phases of probe and source are
equal (in other words, if the radial scalar flow passes through a
$(Q,q)$-marginal stability wall, where $\alpha_q=\alpha$), the
potential $V$ reaches a zero energy minimum (cases (a) and (b) in
the figure). Placed at this radius, the probe does not break more
of the supersymmetry than the background already did, and the
configuration is BPS. The value of $R$ depends on the scalar vevs
$u^a={z^a|}_{r=\infty}$, and is immediately obtained from
(\ref{floweq}) by taking $Q'=q$:
\begin{equation} \label{sugraprobemin}
 R = \left. \frac{l_P \, \langle q,Q \rangle}{2 \im[e^{-i \alpha} Z_q]} \right|_{r=\infty}
 = \left. \frac{\langle q,Q \rangle}{2 \, m \, \sin(\alpha_q-\alpha)}
 \right|_{r=\infty} \, ,
\end{equation}
Of course, $R$ needs to be positive, so a necessary condition for
the existence of such a classical BPS bound state with nonzero
separation (in the probe approximation) is that the DSZ
intersection product $\langle q,Q \rangle$ and
$\sin(\alpha_q-\alpha)_{r=\infty}$ have the same
sign. %\footnote{Though the domain of validity of this classical
%probe/supergravity analysis is, as usual, complementary to the
%domain of validity of classical ($g_s = 0$) stringy D-brane
%analysis, the stability condition obtained here is, for this
%example and after suitable identification of the words
%``subobject'' and ``constituent particle'', essentially identical
%to the $\Pi$-stability criterion \cite{DFR} obtained in the
%D-brane context. This and other parallels will be elucidated and
%further explored in section \ref{sec:qq}.}
Note also that the
phenomenon of decay at marginal stability is a natural, smooth
process in this picture: when we start with a BPS configuration in
a certain vacuum characterized by $u^a=z^a|_{r=\infty}$, and the
$u^a$ are varied to approach a marginal stability wall, the radius
diverges and the BPS bound state decays smoothly into a two
particle state.

Much of the interesting structure emerging here is in fact a
direct consequence of supersymetry, as we will see in the
following.

\subsection{Supersymmetric mechanics of a single probe particle}
\label{sec:oneprobemech}

An (effective) BPS particle in a 3+1 dimensional $\CN=2$ theory
conserves four of the original eight supercharges. Its low energy
dynamics can therefore be expected to be described by a $d=1$,
$\CN=4$ supersymmetric Lagrangian. The degrees of freedom
appearing in this Lagrangian will always at least include the
position coordinate $\sx$ in the noncompact space, together with
its fermionic superpartner, given by a 2-component spinor
$\lambda_\alpha$, $\alpha=1,2$ and its complex conjugate
$\bar{\lambda}^\alpha \equiv (\lambda_\alpha)^*$. Together with an
auxiliary bosonic variable $D$ and a one dimensional connection
$A$, these degrees of freedom form a vector multiplet (or linear
multiplet) on the particle worldline, which can be thought of as
the dimensional reduction of a $d=4$, $\CN=1$ vector multiplet.
This kind of supersymmetric quantum mechanics was first
constructed in \cite{dCR} for a flat target space. It was
generalized to curved spaces in \cite{Sm} as an effective
description of the zero mode dynamics of supersymmetric QED, given
a superfield description in \cite{IS}, and obtained as an
effective theory for chiral SQED in \cite{Sm2}. The superspace
formulation of this model was rediscovered and used to prove a
non-renormalization theorem for $\CN=8$ quantum mechanics in
\cite{DE}.

The supersymmetry transformations are as follows (our conventions
can be found in appendix \ref{conventions}):
\begin{eqnarray}
 \delta A &=& i \, \bar{\lambda} \, \xi - i
 \, \bar{\xi} \,  \lambda \label{susyvar0} \\
 \delta \sx &=& i \, \bar{\lambda} \, \ssigma \, \xi - i
 \, \bar{\xi} \, \ssigma \, \lambda \label{susyvar1} \\
 \delta \lambda &=& \dot{\sx} \cdot \ssigma \, \xi
 + i \, D \, \xi \label{susyvar2} \\
 \delta D &=& - \dot{\bar{\lambda}} \, \xi
  - \bar{\xi} \, \dot{\lambda} \label{susyvar3}
\end{eqnarray}

The particle can have further internal degrees of freedom, but let
us assume for now that these are absent (or that they do not
couple to the position vector multiplet).

Regardless of its specific form, the Lagrangian can be expanded in
powers of velocity:
\begin{equation} \label{velexp}
 L = L^{(1)} + L^{(2)} + \ldots \, ,
\end{equation}
where we assign the following orders to the various quantities
appearing in the Lagrangian:
\begin{equation}
 \CO{(\sx)} = 0, \qquad \CO(\frac{d}{dt}) = 1, \qquad \CO(D) = 1,
 \qquad \CO(\lambda) = 1/2 \, .
\end{equation}
With this assignment, and taking $\CO(\xi)=-1/2$, the
supersymmetry transformations preserve the order, so to have a
supersymmetric total $L$, each individual $L^{(k)}$ has to be
supersymmetric.

In \cite{Sm,DE}, it was shown that a wide class of quadratic
Lagrangians $L^{(2)}$ can be obtained by choosing an arbitrary
``K\"ahler potential'' function $K(\sx)$, giving for the second
order bosonic part of the Lagrangian:
\begin{equation} \label{secondorder}
 L^{(2)}_B = \frac{1}{4} \nabla^2 K(\sx) \, ( \dot{\sx}^2 + D^2 ) \, .
\end{equation}
The possible presence of a first order term $L^{(1)}$ was not
considered in \cite{DE}, but will be of crucial importance in the
present work. In general it will be of the form
\begin{equation} \label{susyconstr:oneparticle}
 L^{(1)} = -U(\sx) \, D + \sA(\sx) \cdot \dot{\sx} + C(\sx) \, \bar{\lambda} \lambda
 + \sC(\sx) \cdot \bar{\lambda} \ssigma \lambda .
\end{equation}
The second term in this expression gives a Lorentz-type force,
while the first term can be thought of as a position-dependent
Fayet-Iliopoulos term. A similar term is well known to play a key
role in two dimensional $\CN=2$ linear sigma-models \cite{WN2}.
The one dimensional counterpart has been considered as well
\cite{dCR,Sm2,HP}, but seems to be less widely known.

Requiring $L^{(1)}$ to be supersymmetric imposes rather strong
constraints on $U$, $\sA$, $C$ and $\sC$. A direct computation
using (\ref{susyvar1})-(\ref{susyvar3}) gives:
\begin{equation}
 \sC = \nabla U = \nabla \times \sA \, , \qquad C=0 \, .
\end{equation}
Allowing a singularity at the origin, the general spherically
symmetric solution to these constraints is, with $r=|\sx|$:
\begin{equation} \label{Ugenform}
 U = \frac{\kappa}{2 r} + \theta \, , \qquad \sA = - \kappa \sA^d \, ,
\end{equation}
with $\theta$ and $\kappa$ constants, and $\sA^d$ a unit Dirac
monopole vector potential as in (\ref{dirpot}). The Dirac
quantization condition requires $\kappa$ to be an integer.
Non-spherically symmetric solutions are also possible of course,
corresponding to dipoles, quadrupoles and so on, but since we
assumed no further internal particle degrees of freedom, we do not
consider those here. In the simplest case, flat space, the
Lagrangian takes the following form:
\begin{equation}
 L= \frac{m}{2} \left( {\dot{\sx}}^2 + {D}^2
 + 2 i \bar{\lambda} \dot{\lambda} \right) - (\frac{\kappa}{2 |\sx|} + \theta) D
 -  \kappa \sA^d \cdot \dot{\sx} - \frac{\kappa \, \sx}{2 |\sx|^3} \cdot
  \bar{\lambda} \ssigma \lambda. \label{LV}
\end{equation}

The physical potential energy for the position $\sx$ is obtained
by eliminating the auxiliary variable $D$ from the Lagrangian. The
precise form of the potential will therefore depend on the form of
each of the terms $L^{(n)}$ in the expansion (\ref{velexp}). For
example if all $L^{(n)}$ with $n \geq 3$ are zero, the potential
is $V(\sx) = U^2/\nabla^2 K$. However, the presence and position
of supersymmetric extrema of the Lagrangian only depends on the
first order term. Indeed, from the supersymmetry variations
(\ref{susyvar1})-(\ref{susyvar3}), it follows that a classical
supersymmetric configuration is given by a time-independent $\sx$
with $\lambda=0$ and $D=0$. For this to be a solution to the
equations of motion, one needs furthermore $\delta L / \delta D =
0$, which is the case if and only if $U = 0$, or with
(\ref{Ugenform}):
\begin{equation}
 r = - \frac{\kappa}{2 \theta} \, .
\end{equation}
Identifying $\kappa = \langle Q,q \rangle$ and $\theta = m
\sin(\alpha_q - \alpha)|_{r=\infty}$, this expression coincides
with (\ref{sugraprobemin}).

To make this match more precise, we would have to construct the
full supersymmetric extention of the supergravity probe Lagrangian
(\ref{Kformula})-(\ref{Mformula}). This is quite complicated in
the fully relativistic version, so we will restrict ourselves here
to the non-relativistic approximation as given by
(\ref{KformulaNR}) and (\ref{VformulaNR}), i.e.\ small velocities
and small phase difference.\footnote{The condition on the phases
cannot be dropped in a consistent nonrelativistic approximation;
for arbitarary phase differences, the supergravity probe
Lagrangian has no supersymmetric extension quadratic in the
velocities.} First observe that in this regime, the extended
bosonic Lagrangian
\begin{equation} \label{extsugraprobeL}
 L_B = \frac{1}{2 l_P} \rho^{-3} |Z_q| (\dot{\sx}^2 + D^2) - \rho^{-1}
 \im[e^{-i\alpha} Z_q] \, D -
 \, \langle Q,q\rangle \, \sA^d \cdot \dot{\sx}
\end{equation}
reduces to the nonrelativistic version of the probe Lagrangian
(\ref{Kformula})-(\ref{Mformula}) after eliminating the auxiliary
field $D$. Note also that the background BPS equations of motion
(\ref{floweq}) imply that the coefficient of $D$ in the linear
term is of the form (\ref{Ugenform}), with $\kappa=\langle Q,q
\rangle$ and $\theta=l_P^{-1} \im[e^{-i \alpha} Z_q]_{\infty} =
m_q \sin(\alpha_q-\alpha) $. The first order terms thus satisfy
the required constraints. Finally, the second order part is of the
form (\ref{secondorder}), so we conclude that the
(nonrelativistic) supergravity probe action indeed has a
supersymmetric extention of the form analyzed in this section,
with the above identifications of $\kappa$ and $\theta$.

\subsection{General multicentered BPS bound states in
supergravity} \label{sec:sugramulti}

The supergravity probe considerations of section
\ref{sec:sugraprobe} can be enhanced to a full analysis of BPS
solutions of the supergavity equations of motion, with an
arbitrary number of centers $\sx_p$, $p=1,\ldots,N$, carrying
arbitrary charges $Q_p$. This was investigated in detail in
\cite{branessugra}. The analysis is technically quite involved
(mainly due to the fact that the corresponding spacetimes are
non-static --- time-independent but with a non-diagonal metric),
but fortunately we will only need some of the conclusions, which
can be stated rather simply. The result which is most relevant for
this paper is that the BPS requirement gives a set of constraints
on the relative positions of the centers, generalizing
(\ref{sugraprobemin}), namely for every center $p$:
\begin{equation} \label{sugraconstr}
 \sum_{q=1}^N \frac{\langle Q_p, Q_q
 \rangle}{|\sx_q - \sx_p|} = \left. 2 \, m_p \, \sin(\alpha_p - \alpha) \right|_{r=\infty} \, ,
\end{equation}
where $m_p = |Z_p|$ is the BPS mass of a particle with charge
$Q_p$ (in the given vacuum), $\alpha_p$ is the phase of $Z_p$,
$\alpha$ the phase of the total central charge $Z=\sum_p Z_p$, and
$\langle \cdot,\cdot\rangle$ denotes as before the DSZ
intersection product. Note that only $N-1$ constraints are
independent, since summing (\ref{sugraconstr}) over all
$p=1,\ldots,N$ gives trivially $0 = 0$.

Another result of interest from \cite{branessugra} is the fact
that these solutions, being non-static, carry an intrinsic angular
momentum, given by the formula
\begin{equation} \label{sugraspin}
 \sJ = \frac{1}{2} \sum_{p<q} \langle Q_p,Q_q \rangle \,
 \frac{\sx_p-\sx_q}{|\sx_p-\sx_q|} \, .
\end{equation}

\subsection{Supersymmetric multi-particle mechanics}

The general one-particle supersymmetric mechanics of section
\ref{sec:oneprobemech} can be generalized to an arbitrary number
$N$ of interacting BPS particles with arbitrary charges. As we saw
in section \ref{sec:sugramulti} in the example of supergravity,
such systems can have a moduli space of classical BPS
configurations, conserving four of the original eight
supercharges. The low energy dynamics of such a system can
therefore be expected to be described by an $\CN=4$ supersymmetric
multi-particle Lagrangian, with degrees of freedom including at
least the position vector multiplets of the different particles
involved. We will again assume that the particles do not have
further internal low energy degrees of freedom (or at least that
the latter don't couple to the position multiplets). In the
context of string theory, where the particles arise as an
effective description of wrapped D-branes, there can nevertheless
still be additional relevant low energy degrees of freedom
corresponding to strings stretching between the wrapped branes.
For well-separated particles (compared to the string scale), such
strings are very massive and can be integrated out safely.
However, when the particles come close to each other, the
stretched strings can become massless or even tachyonic, and the
description in terms of commuting position multiplets only breaks
down.

In this section, we will ignore this phenomenon and study the
general form of $\CN=4$ supersymmetric effective Lagrangians
involving the abelian position multiplets $\left(
\sx_p,D_p,\lambda_p \right)_{p=1}^N$ only. A more refined analysis
of the breakdown of this description and its stringy resolution
will be presented in section \ref{sec:qq}.

The supersymmetry transformations are as in
(\ref{susyvar0})-(\ref{susyvar3}), with the addition of an index
$p$. The general form of the Lagrangian to first order in the
velocities becomes:
\begin{equation} \label{L1form}
 L^{(1)} = \sum_p \left( - U_p \, D_p + \sA_p \cdot \dot{\sx}_p \right) + \sum_{p,q} \left(
 C_{pq} \,
 \bar{\lambda}_p \lambda_q
 + \sC_{pq} \cdot \bar{\lambda}_p \ssigma \lambda_q \right) \, .
\end{equation}

Requiring the Lagrangian to be supersymmetric gives the following
constraints, generalizing (\ref{susyconstr:oneparticle}):
\begin{equation}
 \sC_{pq} = \nabla_p U_q = \nabla_q U_p = \frac{1}{2}
 \left( \nabla_p \times \sA_q + \nabla_q \times \sA_p \right); \qquad C_{pq} = 0
 \, .
\end{equation}
Allowing singularities when two centers coincide, these
constraints are solved by
\begin{equation}
 U_p = \sum_q \frac{\kappa_{pq}}{2 r_{pq}} + \theta_p \, ,
\end{equation}
with $\kappa_{pq} = -\kappa_{qp}$, and $\sA_p$ the vector
potential produced at $\sx_p$ by a set of magnetic monopoles with
charges $\{\kappa_{pq}\}_q$, $q=1,\ldots,N$ at respective
positions $\{\sx_q\}_q$. Plugged into the general form of the
first order Lagrangian (\ref{L1form}), this gives:
\begin{eqnarray}
 L^{(1)}&=&-\sum_p \theta_p D_p - \sum_{p<q} \kappa_{pq} \, L^{int}_{pq}  \label{multiL11} \\
 L^{int}_{pq} &=& \frac{1}{2 r_{pq}} D_{pq} + \sA^d(\sr_{pq}) \cdot \dot{\sr}_{pq}
 + \frac{1}{2 r_{pq}^3} \, \sr_{pq} \cdot \bar{\lambda}_{pq}
\ssigma \lambda_{pq} \, , \label{multiL12}
\end{eqnarray}
where $\kappa_{pq}=-\kappa_{qp}$ are constants,
$\sr_{pq}=\sx_p-\sx_q$, $D_{pq} = D_p - D_q$, $\lambda_{pq} =
\lambda_p - \lambda_q$, and $\sA^d(\sr)$ as in (\ref{dirpot}). The
Dirac quantization condition here is $\kappa_{pq} \in \IZ$.

As in the single particle case, the precise form of the
interaction potential between the particles also depends on the
higher order terms in the Lagrangian, but its supersymmetric
minima, leading to classical BPS bound states, are entirely
determined by $L^{(1)}$. They occur when $U_p$ becomes zero, i.e.\
at positions satisfying for all $p$:
\begin{equation} \label{multieq}
 \sum_q \frac{\kappa_{pq}}{2 r_{pq}} = - \theta_p \, .
\end{equation}
Again, this expression is identical to the supergravity BPS
constraint (\ref{sugraconstr}) with the identifications
\begin{equation} \label{sugraid}
\kappa_{pq} = \langle Q_q,Q_p \rangle \quad \mbox{and} \quad
\theta_p = \im(e^{-i \alpha}Z_p) = m_p \sin(\alpha_p-\alpha).
\end{equation}

\subsection{Special case: the hole or Hall halo.} \label{sec:hallhalo}

\FIGURE[t]{\centerline{\epsfig{file=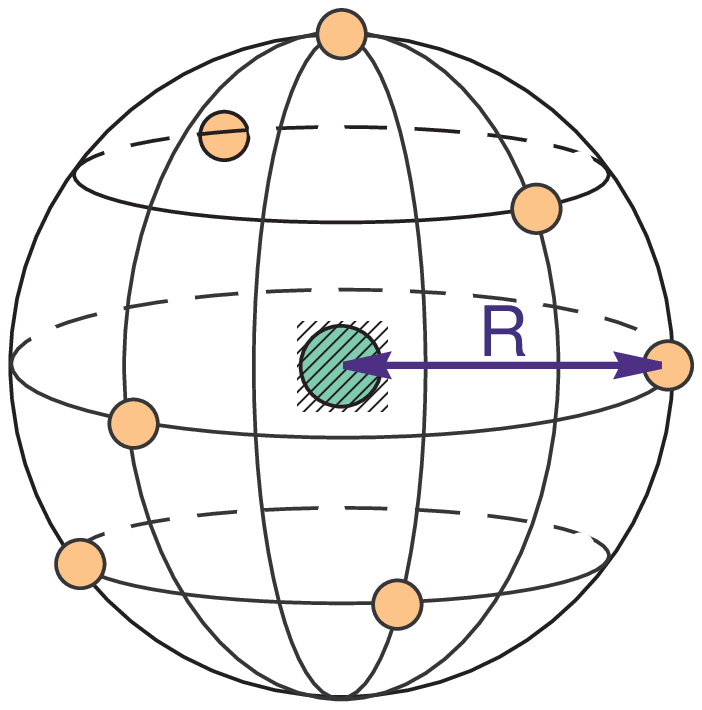,height=6cm}}
\caption{A Hall (or Hole) Halo consisting of a charge $Q$ in the
origin surrounded by $N=7$ charges $q$ on a sphere of radius
$R=-\kappa/2c$, with $\kappa$ units of magnetic flux through the
sphere. }\label{holehalo}}

A simple yet already quite interesting example is the system
consisting of one particle (or black hole) of charge $Q$
interacting with $N$ particles (or black holes) of charge $q$,
with $\kappa \equiv \langle Q,q \rangle \neq 0$. According to
(\ref{multieq}), the classical ground states of this system are
configurations with all $N$ particles of charge $q$ on a sphere of
radius $R=-\kappa/2\theta_q$ around the charge $Q$ particle. In
the supergravity context $\theta_q=l_P^{-1} \left. \im(\bar{Z}_Q
Z_q)/|Z_Q+N Z_q| \, \right|_{\infty}=\mu \sin(\alpha_q-\alpha_Q)$,
where $\mu$ is the ``reduced mass'', $\mu=m_Q m_q / m_{tot}$. One
could call such a configuration a ``Hole Halo'', as illustrated in
fig.\ \ref{holehalo}. From equation (\ref{multiL12}), we
furthermore see that the particles are moving in a uniform
magnetic\footnote{By ``magnetic'' we mean here magnetic relative
to the charges on the sphere. With respect to a fixed chosen
charge basis, the field is not necessarily purely magnetic;
depending on the choice of basis, it could even be purely electric
in that sense.} field with $\kappa$ units of flux through the
sphere. This is the typical setup used to study the quantum Hall
effect on a sphere (see for example \cite{qhall}). One could
therefore equally well call such a system a Hall Halo. Another
system designed to reproduce the spherical quantum Hall effect was
discussed in the context of string theory in \cite{giantgraviton}.
The quantum Hall halo considered here seems to free of some of the
undesired features of this other system \cite{GR}, though we did
not analyze this in detail.

%%%%%%%%%%%%%%%%%%%%%%%%%%%%%%%%%%%%%%%%%%%%%%%%%%%%%%%%%%%%%%%%%%%%%%%%
\section{BPS D-branes and classical quiver mechanics}\label{sec:micro}
\setcounter{equation}{0}
%%%%%%%%%%%%%%%%%%%%%%%%%%%%%%%%%%%%%%%%%%%%%%%%%%%%%%%%%%%%%%%%%%%%%%%%

In string theory, BPS states can often be analyzed perturbatively
by describing them as D-branes, i.e.\ subspaces on which open
strings can end (or their generalizations as CFT boundary states).
This description of BPS states as infinitely thin superposed
objects, with no backreaction on the ambient space, is exact in
the limit of vanishing string coupling constant, $g_s = 0$. When
the string coupling constant is turned on, the D-branes become
dynamical objects of finite width, interacting with the ambient
space, and in suitable regimes (typically at large $g_s \times
\mbox{(number of branes)}$, outside the domain of validity of open
string perturbation theory), they are believed to become
well-described by the solitonic $p$-brane solutions of
supergravity.

In the framework of type II string theory compactified on a
Calabi-Yau manifold, we thus expect the bound states considered in
section \ref{sec:particles} to have a corresponding description as
wrapped D-branes, which becomes accurate when $g_s \to 0$. At
first sight, the fact that those multicentered, molecule-like
bound states have a dual description as a wrapped D-brane
localized at a single point in the noncompact space may seem a bit
odd. However, note that all length scales appearing in section
\ref{sec:particles} are proportional to the four dimensional
Planck length $l_P$, related to the string length $l_s = \sqrt{2
\pi \alpha'}$ as
\begin{equation} \label{planckstringrelation}
 l_P = g_s l_s/\sqrt{v} \, ,
\end{equation}
where $v=2V/\pi^2l_s^6$, with $V$ the volume of the Calabi-Yau.
Thus, if we take $g_s \to 0$ while keeping all other parameters
fixed, the equilibrium radii such as (\ref{sugraprobemin}) become
vanishingly small compared to the string length. The naive
particle picture breaks down in this limit, as in particular open
strings stretching between the different branes can become
tachyonic, leading to the decay of the multicentered configuration
into a single-centered wrapped brane. Conversely, this also
indicates that quantum effects (i.e.\ open string loops) should
produce rather drastic qualitative effects in order to match the
two, allowing to go smoothly from single centered wrapped branes
into multicentered configurations. Details of this D-brane bound
state metamorphosis will be analyzed in section \ref{sec:qq}.

The classical ($g_s = 0$) D-brane description of BPS states in
Calabi-Yau compactifications has been studied extensively (an
incomplete list of references is \cite{DMquiv}-\cite{Dcat},
\cite{joyce}-\cite{DGJT}). One of the results emerging from this
work is that the picture of BPS D-branes as classical submanifolds
(possibly carrying certain vector bundles), valid in the large
radius limit, needs to be modified for generic moduli of the
Calabi-Yau. The more general picture is that of a D-brane as an
object in a certain category, with massless fermionic strings
playing the role of morphisms between the objects. The full story
requires quite a bit of algebraic geometry, but fortunately, in
many cases, the low energy D-brane dynamics can simply be
described by a $d=4$, $\CN=1$ quiver gauge theory dimensionally
reduced to the effective particle worldline, and we can forget
about the underlying geometrical structure. This is well known in
the case of orbifold constructions \cite{DMquiv}. Another (though
overlapping) class of examples is given by collections of
``parton'' D-branes with nearly equal phases. Parton D-branes
\cite{fiol} are D-branes which can be considered elementary (at
the given point in moduli space), in the sense that they come in
the smallest massive BPS supermultiplets of the theory
(hypermultiplets for an $\CN=2$ theory), and that other D-branes
can be constructed as their bound states. This typically
corresponds to D-branes that can become massless (in four
dimensional Planck units) at some point in Calabi-Yau moduli
space, such as a type IIB D3-brane wrapped around an $S^3$ cycle
vanishing at a conifold point.\footnote{To make this statement
more precise, we would have to use the category construction of
\cite{Dcat}. Since we only want to give some intuition in how the
quiver description arises, and are ultimately only interested in
the resulting (dimensionally reduced) gauge theory, we will not
get into this more deeply here.}

\subsection{An example} \label{sec:example}

\FIGURE[t]{\centerline{\epsfig{file=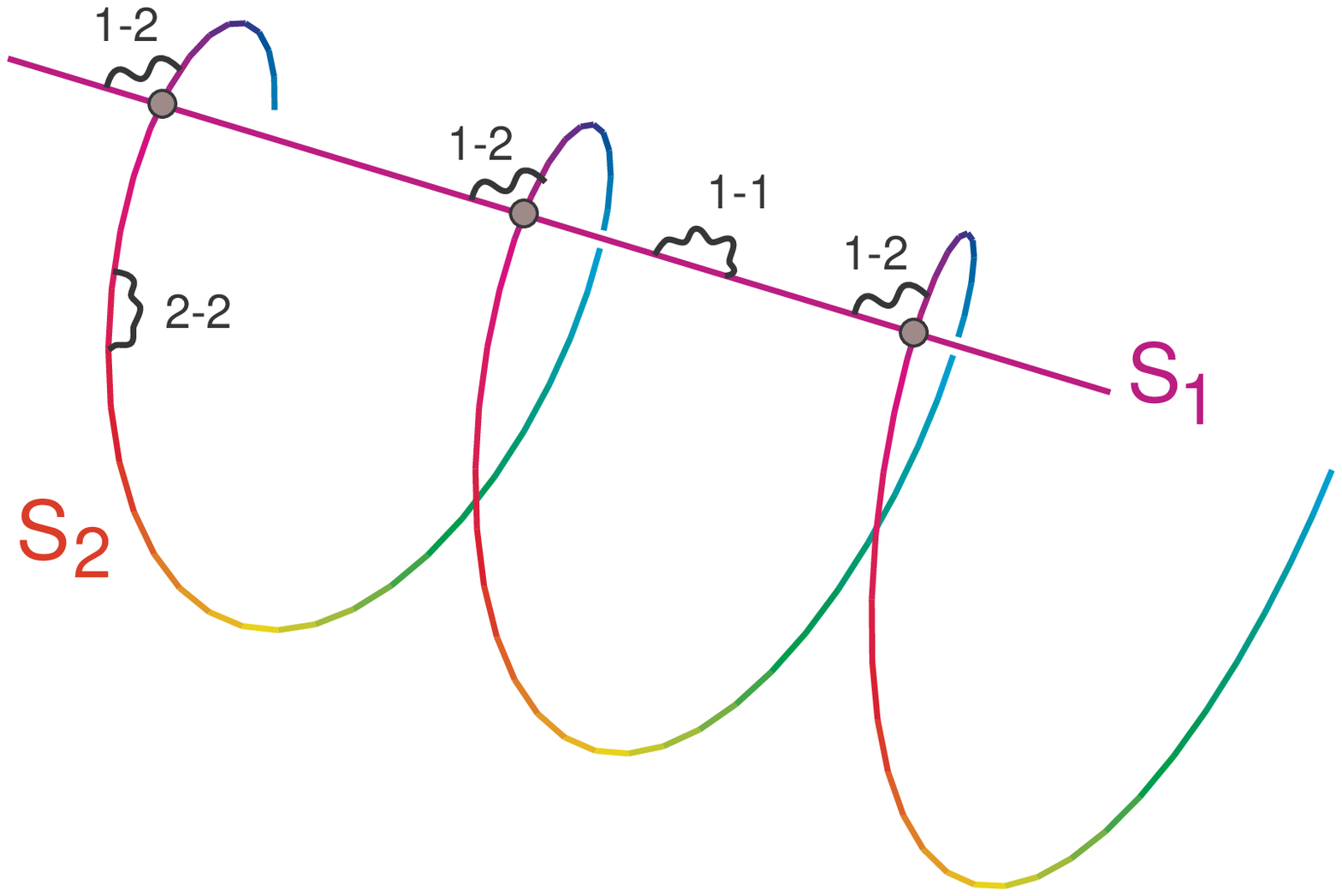,height=6cm}}
\caption{One dimensional caricature of the intersecting branes
$S_1$ and $S_2$. The various light open string modes are
indicated.}\label{intersecting}}

We start by considering a simple intuitive model in type IIB, in a
regime where the classical geometric picture of D-branes as
minimal volume manifolds is accurate, or at least where thinking
of these branes geometrically gives the right results for our
purposes.

Let $S_1$ and $S_2$ be two parton D3-branes with the topology of a
3-sphere, wrapped around two distinct cycles in a Calabi-Yau $X$.
We assume $S^3$ topology for this example because
$H_1(S^3,\IZ)=0$, implying \cite{joyce,hitchin,SYZ} that the
individual branes are \emph{rigid}, i.e.\ they have no moduli of
their own.\footnote{Examples of non-rigid parton D-branes are
given by flat D3 branes on $T^6$; they are partons because they
come in the smallest possible $\CN=8$ massive BPS multiplet, they
can be used to build more complicated branes \cite{slags} and
their mass (in four dimensional Planck units) vanishes at
appropriate large complex structure points, but they are not rigid
since they have translation moduli.} Assume furthermore that $S_1$
and $S_2$ intersect transversally in $\kappa$ points, with all
intersections positive, so the geometric intersection product
$\langle S_1,S_2 \rangle = \kappa > 0$. The situation is sketched
in fig.\ \ref{intersecting}. The central charges of the branes are
given by $Z_p = \int_{S_p} \Omega$, with $\Omega$ the holomorphic
3-form on $X$, normalized so $i \int \Omega \wedge \bar{\Omega} =
1$. The central charge phases are denoted as $\alpha_p = \arg
Z_p$, $\alpha=\arg Z$. The combined brane system is supersymmetric
(BPS) if $\alpha_1=\alpha_2$.

There are three different kinds of open string modes: starting and
ending on $S_1$, starting and ending on $S_2$, and stretching
between $S_1$ and $S_2$. The massless $1-1$ and $2-2$ string modes
correspond to changes in the positions $\sx_p$ of the branes in
the noncompact space, and possibly to further ``internal''
susy-preserving deformations of the individual branes. The latter
are absent though in the case at hand, since we assumed the $S_p$
rigid, leaving only the position modes. They come in two $d=1$,
$\CN=4$ vector multiplets (dimensionally reduced $d=4$, $\CN=1$
$U(1)$ vector multiplets), which we will denote as in the previous
section by $(A_p,\sx_p,D_p,\lambda_p)$, $p=1,2$. The functions
$A_p$ are the one-dimensional $U(1)$ connections.

Assuming one can locally use the branes at angles setup of
\cite{BDL} to compute the light $1-2$ open string spectrum, one
finds that in the case of coincident positions and phases
($\sx_1=\sx_2$ and $\alpha_1 = \alpha_2$), there are $\kappa$
massless modes, corresponding to stretched strings localized at
the intersection points. These modes come in charged chiral
multiplets, with charge $(-1,1)$ under the $U(1) \times U(1)$
D-brane gauge group. The sign of the charge and the chirality are
determined by the sign of the intersection. We denote these chiral
multiplets by $(\phi^a,F^a,\psi^a)$, $a=1,\ldots,\kappa$, where
$\phi^a$ is a complex scalar, $\psi^a$ a 2-component spinor, and
$F^a$ an auxiliary complex scalar.

When the branes are separated in the noncompact space,
supersymmetry is preserved but the chiral multiplets become
massive with mass $m_C=|\sx_2-\sx_1|$ (in units such that $l_s
\equiv 1$, which we will use from now on). On the other hand when
$\alpha_1 \neq \alpha_2$, supersymmetry is broken and bose-fermi
degeneracy in the chiral multiplets is lifted: while the mass of
the fermionic chiral modes $\psi^a$ remains unchanged, the mass of
the bosonic modes $\phi^a$ is shifted to ${m_{\phi}}^2 =
(\sx_2-\sx_1)^2 + \alpha_2-\alpha_1$. For sufficiently small
separation and $\alpha_2<\alpha_1$, these modes become tachyonic.
Tachyon condensation corresponds in this case to the formation of
a classical bound state, a single wrapped D-brane $S$ produced by
putting $\sx_1=\sx_2$ and deforming the intersections to throats
smoothly connecting $S_1$ and $S_2$. This breaks (classically) the
$U(1) \times U(1)$ gauge symmetry to $U(1)$, and the corresponding
$\kappa-1$ massless Goldstone bosons provide the deformation
moduli of $S$, consistent with the fact that $b^1(S=S_1 \#
S_2)=\kappa-1$. Note that the condition $\alpha_1-\alpha_2>0$ for
bound state formation here is the same as the one we found in the
context of supergravity (remember we took $\kappa = \langle
S_1,S_2 \rangle > 0$).

To keep these chiral multiplets much lighter than the infinite
tower of excited open string modes, so they deliver the dominant
contribution to the inter-brane interaction at low energies, we
take $|\alpha_1-\alpha_2| \ll 1$ and $|\sx_1-\sx_2| \ll l_s$. Then
the following classical $d=1$, $\CN=4$ Lagrangian $L=L_V+L_C$,
obtained by dimensional reduction from the $\CN=1$, $d=4$ gauge
theory with the field content outlined above, describes the low
energy dynamics of these branes:
\begin{eqnarray}
 L_V &=& \frac{m_p}{2} \left( {\dot{\sx}_p}^2 + {D_p}^2
 + 2 i \bar{\lambda_p} \dot{\lambda}_p \right) - \theta_p D_p
 \label{LVquiv}
 \\
 L_C &=& |\CD_t \phi^a|^2
 - \left( ({\sx_2-\sx_1})^2 + D_2 - D_1 \right) |\phi^a|^2
 + |F^a|^2
 + i \, \bar{\psi^a} \CD_t \psi^a \nonumber \\
 &&
 - \bar{\psi}^a \, (\sx_2-\sx_1) \cdot \ssigma \, \psi^a
 - i \sqrt{2} (\bar{\phi}^a \psi^a \epsilon (\lambda_2-\lambda_1) -
 (\bar{\lambda}_2 - \bar{\lambda}_1) \epsilon \bar{\psi}^a) \phi^a \, ,
\end{eqnarray}
where summation over $p=1,2$ and $a=1,\ldots,k$ is understood, the
covariant derivative $\CD_t \phi \equiv \left( \partial_t +
i(A_2-A_1) \right) \phi$, $m_p=|Z_p|/l_{P}$, the $\theta_p$ are
Fayet-Iliopoulos parameters, and we have put $l_s=1$. We will see
below that we have to take $\theta_p \equiv m_p(\alpha_p -
\alpha_0)$, with $\alpha_0 \equiv (m_1 \alpha_1 + m_2
\alpha_2)/(m_1+m_2)$, to match the string masses and to have zero
energy for supersymmetric configurations.\footnote{This is a
convention. The value of the rest energy does not influence the
particle dynamics.} The supersymmetry variations under which this
action is invariant can be found in appendix \ref{app:susy}. This
Lagrangian can also be considered to be the dimensional reduction
of a two-dimensional linear sigma model \cite{WN2}. These models
have been analyzed very extensively in the literature, and since
the \emph{classical} features of the $d=2$ and the $d=1$ versions
are essentially identical, the following discussion is in essence
merely a review of well known facts. It is useful though to have
things explicit for subsequent sections.

The Lagrangian can be split in a center of mass part and a
relative part. Denoting the center of mass variables by $\sx_0
\equiv (m_1 \sx_1 + m_2 \sx_2)/(m_1+m_2)$ and so on, the center of
mass Lagrangian is simply
\begin{equation}\label{comdof}
  L_0 = \frac{m_1+m_2}{2} \left( {\dot{\sx_0}}^2 + {D_0}^2
 + 2 i \bar{\lambda}_0  \dot{\lambda}_0 \right).
\end{equation}
There is no FI-term, hence no rest energy, because
$\theta_1+\theta_2=0$.

Denoting the relative variables by $\sx = \sx_2 - \sx_1$ and so
on, the relative part of the Lagrangian $L_{rel} = L_{rel,V} +
L_{rel,C}$ becomes
\begin{eqnarray}
 L_{rel,V} &=& \frac{\mu}{2} \left( {\dot{\sx}}^2 + {D}^2
 + 2 i \bar{\lambda} \dot{\lambda} \right) - \theta D \label{relLV} \\
 L_{rel,C} &=&  |\CD_t \phi^a|^2 - \left( \sx^2 + D \right) |\phi^a|^2 +
 |F^a|^2+ i \, \bar{\psi^a} \CD_t \psi^a \nonumber \\
 &&
 - \bar{\psi^a} \, \sx \cdot \ssigma \, \psi^a
 - i \sqrt{2} (\bar{\phi}^a \psi^a \epsilon \lambda -
 \bar{\lambda} \epsilon \bar{\psi}^a \phi^a) \, , \label{relLC}
\end{eqnarray}
where the reduced mass $\mu=m_1 m_2/(m_1+m_2)$ and $\theta
 = -\theta_1 = \theta_2 = \mu (\alpha_2-\alpha_1)$. The gauge group is
the relative $U(1)$ between the branes. Eliminating the auxiliary
fields $D$ and $F$ yields the following potential:
\begin{equation} \label{higgspotential}
 V(\phi,\sx) = \frac{1}{2 \mu} \left( |\phi|^2 + \theta \right)^2 +
 |\sx|^2 | \phi |^2 \, ,
\end{equation}
with $|\phi|^2 \equiv \sum_a |\phi^a|^2$. In particular this gives
for the mass\footnote{In the context of particle mechanics, the
``mass'' of this mode is actually its oscillator frequency, but we
will use the term mass as well, hoping that this will not lead to
confusion.} of the $\phi$-modes ${m_\phi}^2 = |\sx|^2 + \theta/\mu
= |\sx|^2 + \alpha_2-\alpha_1$, correctly matching the stretched
bosonic string masses as given earlier. As a further check, note
that at $\phi=0$, the potential energy equals $\mu
(\alpha_2-\alpha_1)^2/2 \approx |Z_1| + |Z_2| - |Z_1 + Z_2|$, in
agreement with the Born-Infeld D-brane action if $\phi=0$ is
interpreted as having disconnected branes.

The moduli space of classical ground states of this system is
given by the local minima of $V$. If $\theta<0$, there are two
branches: one (the ``Coulomb'' branch) consisting of
configurations with $\phi=0$ and $\sx^2>\alpha_1-\alpha_2$, and
another one (the ``Higgs'' branch) consisting of configurations
with $\sx=0$ and $|\phi|^2=-\theta$ modulo the $U(1)$ gauge group.
Configurations with $\phi=0$ and $\sx^2<\alpha_1-\alpha_2$ are
unstable; the $\phi$-modes become tachyonic and tend to
``condense'' into a Higgs branch ground state. Note that (for
$\theta<0$) the Higgs branch is $\CP^{\kappa-1}$, and the Coulomb
branch $\IR^3$ with a ball removed. If $\theta > 0$, there is no
Higgs branch, and the Coulomb branch is $\IR^3$.

A ground state is supersymmetric if $D=0$ or equivalently $V=0$,
as can be seen directly from the supersymmetry transformation
rules (\ref{rule1})-(\ref{finalrule}). Thus at the classical
level, only the Higgs branch can provide supersymmetric ground
states, unless $\theta=0$; then it is the Coulomb branch. All this
matches the string theory features discussed earlier.

We emphasize that these considerations are all classical. Quantum
effects drastically alter this picture, as we will see in section
\ref{sec:qq}.

\subsection{General model and quiver mathematics} \label{quivmath}

\FIGURE[t]{\centerline{\epsfig{file=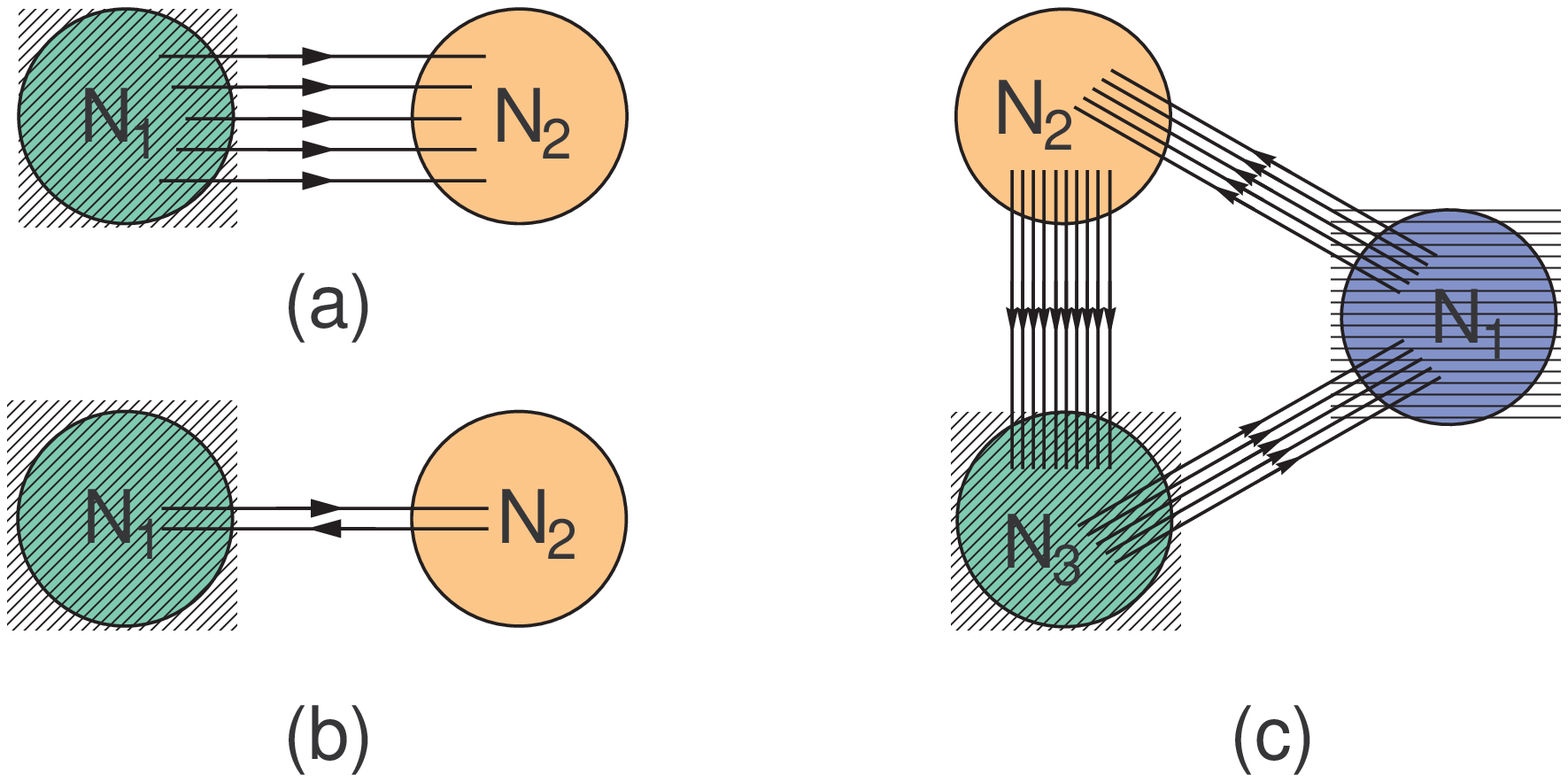,height=6cm}}
\caption{Some examples of quiver diagrams. The $N_v$ indicate the
dimensions of the vector spaces corresponding to the nodes. The
quivers (a) and (c) appear in the description of certain D-branes
on the quintic \cite{DD,Dcat,DGJT}. For example case (a)
represents a bound state of $N_1$ pure $D6$-branes with $N_2$ $D6$
branes carrying the bundle ${\cal O}(-1)$. In the mirror picture
the corresponding branes are D3-branes wrapped around two
different cycles vanishing at the conifold point copies $\psi=1$
resp.\ $\psi=e^{2 \pi i/5}$ (in the notation of \cite{DGR}). The
two branes have intersection product equal to five, hence the five
arrows.}\label{quivers}}

The model discussed in the previous subsection, describing the low
energy dynamics of two rigid parton D-branes $S_1$ and $S_2$ with
nearly coincident positions and phases, and with $\kappa$ light
chiral open strings between them, can be generalized in various
ways. For example, one could increase the number of both types of
branes, say to $N_p$ branes of type $S_p$ ($p=1,2$). With $N_1=1$,
this gives the D-brane counterpart of the Hole (Hall) Halo of
section \ref{sec:hallhalo}. The position coordinates of the stack
of type $S_p$ branes now become $N_p \times N_p$ hermitian
matrices, with the off-diagonal elements representing strings
stretching between different copies of the same $S_p$, while the
strings stretching from an $S_1$ type brane to an $S_2$ type brane
are now represented by $\kappa$ $N_2 \times N_1$ complex matrices
transforming in the $(\boldsymbol{\bar{N}_1},\boldsymbol{N_2})$ of
the $U(N_1) \times U(N_2)$ gauge group of the brane system. A
further generalization is to include branes of other types. For
simplicity we will always assume that the constituent branes $S_p$
are rigid partons. In all these cases, for small separations and
phase differences, the low energy dynamics is expected to be given
by an $\CN=1$, $d=4$ quiver gauge theory reduced to the effective
particle worldline, with field content given by a quiver diagram.

Let us first recall some definitions from the mathematical theory
of quivers \cite{king,nakajima} (a more elaborate summary can be
found e.g.\ in \cite{DFR2}).\footnote{Some of the definitions we
recall here, in particular those concerning $\theta$-stability,
are only given to make contact with the mathematical results of
\cite{reineke}, which we will apply to solve ground state counting
problems. They are however not strictly necessary to understand
the main idea of this paper.} A \emph{quiver diagram} $Q$ is an
oriented graph, consisting of nodes (or vertices) $v \in V$ and
arrows $a \in A$. Some examples are shown in fig.\ \ref{quivers}.
A ($\IC$-)\emph{representation} $R=(X,\phi)$ of a quiver is given
by a set of vector spaces $X_v=\IC^{N_v}$ associated to the nodes
$v \in V$, and a set of linear maps $\phi^a:V_v \to V_w$
associated to the arrows $a:v \to w \in A$. Such linear maps can
be represented by $N_w \times N_v$ complex matrices. The vector
$\sN \equiv (N_v)_{v \in V}$ is called the \emph{dimension vector}
of the quiver representation. A \emph{morphism} between two
representations $(\tilde{X},\tilde{\phi})$ and $(X,\phi)$ is a set
of linear maps $(T_v)_{v \in V}$, with $T_v:\tilde{X}_v \to X_v$,
such that $\phi^a T_v = T_w \tilde{\phi}^a$ for all arrows $a:v
\to w$. A representation $(\tilde{X},\tilde{\phi})$ is called a
\emph{subrepresentation} of $(X,\phi)$ if there exist an injective
morphism from the former to the latter. Two representations
$(X,\phi)$ and $(X,\tilde{\phi})$ are considered equivalent if
they are isomorphic, in other words if there exist a
``complexified gauge transformation'' $(g_v)_{v \in V} \in
\prod_{v \in V} GL(N_v) \equiv G_{\IC}$ such that $\tilde{\phi}^a
= g_w \phi^a g_v^{-1}$ for all arrows $a:v \to w \in A$. We will
denote the subgroup of \emph{unitary} isomorphisms (``ordinary
gauge transformations'') by $G$, that is, $G \equiv \prod_{v \in
V} U(N_v)$.

An important concept in the study of quivers is the notion of
$\theta$-\emph{(semi-)stability} \cite{king}. Let $(\theta_v)_{v
\in V}$ a set of real numbers associated to the nodes of $Q$,
satisfying $\theta(\sN) \equiv \sum_v N_v \theta_v = 0$ for a
given dimension vector $\sN$. Then a representation $R$ is called
$\theta$-stable if every proper subrepresentation $\tilde{R}$ of
$R$ satisfies $\theta(\tilde{\sN}) = \sum_v \tilde{N}_v \theta_v <
0$. Semi-stability is the same with the ``$<$'' replaced by
``$\leq$''. The moduli space of $\theta$-semi-stable isomorphism
classes of representations with dimension vector $\sN$, denoted
$\CM^{ss}(Q,\sN,\theta)$, is a projective variety, and the
subspace obtained by considering only stable representations,
denoted $\CM^s(Q,\sN,\theta)$ is a smooth open subvariety
\cite{king}. The connection with physics will be made below
through a theorem by King \cite{king}, which states that if a
representation is stable, it is equivalent to exactly one solution
of
\begin{equation} \label{Dflatness}
 \sum_{a:v \to *} {\phi^a}^\dagger \phi^a - \sum_{a:* \to v} \phi^a
 {\phi^a}^\dagger = \theta_v \, \one_{N_v} \qquad \forall v \in V.
\end{equation}
modulo the group of unitary gauge transformations $G$. A
semi-stable representation is equivalent to \emph{at most} one
solution of these equations. A non-semi-stable representation on
the other hand is never equivalent to a solution. If we denote the
moduli space of solutions to this equation modulo $G$ by
$\CM(Q,\sN,\theta)$, we thus have $\CM^s \subseteq \CM \subseteq
\CM^{ss}$, and the three spaces do not necessarily coincide. This
discrepancy is mathematically inconvenient, but one can remove it
by introducing the notion of S-equivalence \cite{king,FM}, or, as
we will do, by simply assuming the dimension vector and the
$\theta_v$ to be sufficiently generic, such that all semi-stable
representations are automatically stable. More precisely, as can
be easily verified, this is achieved by taking
$\mbox{g.c.d.}\{N_v\}_{v \in V} = 1$ and the $\theta_v$ linearly
independent over $\IQ$ except for the relation $\sum_v N_v
\theta_v = 0$.

\subsection{Quiver physics}

We now turn to the physical interpretation of all this. A quiver
diagram with given dimension vector $\sN$ is associated to the
field content of an $\CN=1$, $d=4$ gauge theory (or, for our
purposes, its dimensional reduction, to $d=1$) in a natural way:
each node $v$ corresponds to a vector multiplet for gauge group
$U(N_v)$, and each arrow $a:v \to w$ corresponds to a chiral
multiplet transforming in the
$(\boldsymbol{\bar{N}_v},\boldsymbol{N_w})$ of $U(N_v) \times
U(N_w)$. So a quiver representation as defined above is nothing
but a particular configuration of the chiral multiplet scalars in
a particular gauge theory, and gauge transformations correspond to
the group $G$ of unitary isomorphisms. The explicit Lagrangian for
a given quiver, together with the relevant supersymmetry
transformations, is given in appendix \ref{app:genquiv}.

The example considered in section \ref{sec:example} thus
corresponds to a quiver with two nodes and $\kappa$ arrows from
the first one to the second, and dimension vector $\sN=(1,1)$. The
generalization to $N_1$ branes of type $S_1$ and $N_2$ branes of
type $S_2$ is represented by the same quiver with dimension vector
$(N_1,N_2)$. This is illustrated in fig.\ \ref{quivers}a for
$\kappa=5$. More generally one can think of a quiver with
dimension vector $\sN$ as arising from a collection of
D3-branes\footnote{Of course the quiver can also arise from a
completely different geometric setup, e.g.\ from even dimensional
branes with bundles in IIA, or from orbifold constructions or
abstract conformal field theory considerations
--- the D3-brane construction just gives a particularly catchy
geometrical picture.} wrapped $N_v$ times around supersymmetric,
transversally intersecting parton 3-cycles $S_v$ with nearly equal
phase angles $\alpha_v$,\footnote{Note that generically, this can
only be realized physically for some value of the closed string
moduli if the number of different cycles is at most one more than
the real dimension of the vector multiplet moduli space.} where a
positive intersection point between $S_v$ and $S_w$ corresponds to
an arrow from $v$ to $w$ and a negative one to an arrow in the
opposite direction. The arrows can be thought of as the light
stretched strings localized near the intersection points.

Of course, the field content alone does not fix the gauge theory.
One also needs to specify the Fayet-Iliopoulos coefficients
$(\theta_v)_{v \in V}$, and the superpotential $W(\phi)$, which is
a holomorphic function of the chiral fields $\phi^a$, $a \in A$.
When the quiver does not have closed loops (i.e.\ if one cannot
return to the same place by following the arrows), the requirement
of gauge invariance forbids a superpotential. When closed loops
are present (like in fig.\ \ref{quivers}b and \ref{quivers}c) ,
this is no longer the case, and a superpotential will generically
appear. Determining this superpotential from the D-brane data is a
difficult problem in general, though significant progress has been
made \cite{DGJT}. On the other hand, determining the FI-parameters
is easy. As in our basic example, they are obtained by comparing
the relevant parameters in the quiver Lagrangian (see appendix
\ref{app:genquiv}) with the known masses of bosonic strings
stretched between the different branes. This gives
\begin{equation} \label{FIgen}
 \theta_v = m_v (\alpha_v - \alpha_0),
\end{equation}
where $\alpha_v = \arg Z_v$, $Z_v$ being the central charge of the
brane labeled by $v$, and $\alpha_0 \equiv \sum_v N_v m_v
\alpha_v/\sum_v N_v m_v$ (determined by requiring $\sum_v N_v
\theta_v = 0$, equivalent to vanishing energy for supersymmetric
configurations). Note that since all phases are nearly equal, we
have $\alpha_0 \approx \alpha = \arg Z$ and $\theta_v \approx
\im(e^{-i\alpha} Z_v).$

As can be seen from the Lagrangian and the supersymmetry
transformations in appendix \ref{app:genquiv}, for generic
FI-parameters, the classical ground states are given by commuting
and coincident positions $X$, and values of $\phi$ which satisfy
simultaneously the F-flatness conditions, i.e.\ $F^a=\partial_a
W=0$, and the D-flatness conditions, i.e.\ $D_v=0$. The latter
happen to be precisely the equations (\ref{Dflatness}), as can be
seen from the Lagrangian in appendix \ref{app:genquiv}. So if
$W=0$, the classical moduli space is $\CM(Q,\sN,\theta)$. As
discussed earlier, for sufficiently generic $\theta$, this space
coincides with $\CM^{ss}$ and $\CM^s$, guaranteeing that it is a
smooth projective variety.

As an example, consider the quiver with two nodes and $\kappa$
arrows, and dimension vector $\sN=(1,N_2)$. The case $N_2=1$ is
the example of section \ref{sec:example}. For general $N_2$, we
have the quiver corresponding in particle content to the Hall Halo
of section \ref{sec:hallhalo}. The gauge group is $U(1) \times
U(N_2)$. The scalar fields are grouped in $\kappa$ row vectors
$\phi^a$ with $N_2$ entries each, transforming in the fundamental
of $U(N_2)$ and with charge $-1$ under the first $U(1)$. The
FI-parameters are $\theta_2=-
\theta_1/N_2=\mu(\alpha_2-\alpha_1)$, with reduced mass $\mu=m_1
m_2 / (m_1 + N_2 m_2)$. There are no closed loops, so there is no
superpotential. The D-flatness conditions (\ref{Dflatness}) are
\begin{eqnarray}
 \sum_{a,n} \bar{\phi}^a_n \phi^a_n &=& \mu (\alpha_1 - \alpha_2)
 N_2
 \, \label{Dflatex1} \\
 \sum_a \phi^a_n \bar{\phi}^a_m &=& \mu (\alpha_1 - \alpha_2)
 \delta_{nm}
 \label{Dflatex2} \, .
\end{eqnarray}
The first equation follows from the second, corresponding to the
fact that the sum over all $v$ of (\ref{Dflatness}) trivially
leads to $0=0$. If $\alpha_2<\alpha_1$ and $N_2 \leq \kappa$, the
moduli space $\CM$ of solutions to (\ref{Dflatex2}) modulo gauge
transformations is nonempty. It is essentially (up to a
normalization factor) the space of all possible orthonormal
$N_2$-tuples $(\phi_n)_n$ of vectors in $\IC^{\kappa}$, modulo
$U(N_2)$-rotations. In other words, it is the space of all
$N_2$-dimensional planes in $\IC^\kappa$, also known as the
\emph{Grassmannian} $Gr(N_2,\kappa)$. Note that $\dim_{\IC}
Gr(N_2,\kappa) = N_2 (\kappa-N_2)$, in agreement with
straightforward counting of solutions to (\ref{Dflatex2}) minus
the number of gauge symmetries broken by a generic solution. In
the case $N_2=1$, we get $\CM=Gr(1,\kappa)=\CP^{\kappa-1}$,
reproducing the result of section \ref{sec:example}. All this
could alternatively be analyzed in terms of $\theta$-stability,
but we will not do this here.

The main advantage though of the description of quiver moduli
spaces in the language of $\theta$-stability is that it makes it
possible to explore their properties in a systematic, algebraic
way. For applications to D-brane physics, a particularly important
property of these spaces is their cohomology, since cohomology
classes can be identified with supersymmetric ground states, and
betti numbers therefore with various ground state degeneracies. As
the topology of these moduli spaces is typically extremely rich,
computing their betti numbers is a difficult mathematical problem.
Nevertheless, recently, Reineke succeeded in constructing an
explicit formula for the generating function of these numbers
(also known as the Poincar\'e polynomial), for arbitrary quivers
without closed loops \cite{reineke}. We will give some
applications of this powerful result in section \ref{sec:LLL}.

%%%%%%%%%%%%%%%%%%%%%%%%%%%%%%%%%%%%%%%%%%%%%%%%%%%%%%%%%%%%%%%%%%%%%%%%
\section{Quiver quantum mechanics}\label{sec:qq}
\setcounter{equation}{0}
%%%%%%%%%%%%%%%%%%%%%%%%%%%%%%%%%%%%%%%%%%%%%%%%%%%%%%%%%%%%%%%%%%%%%%%%

\subsection{Relating the two pictures} \label{sec:relating}

We have developed two rather different classical pictures of
supposedly the same BPS bound state of branes, on the one hand the
picture of a set of particles at equilibrium separations from each
other and on the other hand the picture of a fusion of D-branes
(with nearly equal phases) at a single point of space. Upon
(stringy) quantization, we should be able to connect the two by
continuous variation of the coupling constants. The aim of this
section is to find out how exactly this comes about.

Let us first go back to the example of section \ref{sec:example}.
At first sight, also the quantum mechanics traditionally
associated to the two pictures looks very different: the first one
gives a purely spatial Schr\"odinger equation of an
electron-monopole pair bound to each other by a potential with
finite distance minimum, whereas the second one gives rise to
quantum mechanics on the quiver moduli space $\CP^{\kappa-1}$,
with a trivially flat Coulomb branch.

The mystery dissipates in a way similar to (but not quite the same
as) the resolution of some puzzles in $\CN=2$, $d=2$ sigma models
\cite{WN2,W3}. Note that the full quiver quantum mechanics
involves both vector and chiral modes. If we take the constituent
branes sufficiently far apart in space, the charged chiral modes
become massive and can be integrated out, giving a contribution to
the low energy effective Lagrangian of the (still massless)
position multiplets alone. Given the non-renormalization theorem
of section \ref{sec:particles}, it seems very plausible that this
reproduces exactly the right first order part $L^{(1)}$ of the
Lagrangian to make the match between the two pictures. Here we
will verify this explicitly.

The Lagrangian (\ref{relLC}) for the chiral modes is quadratic, so
integrating out can be done exactly, either by Wick rotating and
doing Gaussian integrals, or by computing one loop diagrams. With
$r \equiv |\sx|$, the mass (or oscillator frequency) of the
bosonic chiral mode $\phi^a$ is $\sqrt{r^2 + D}$, the mass of the
fermions $\psi^a$ is $r$, and in total there are $\kappa$ such
pairs $(\phi^a,\psi^a)$. The resulting bosonic effective
Lagrangian for constant $\sx$ and $D$ is
\begin{eqnarray}
 L_{B,eff,const.} &=& \frac{\mu}{2} D^2 - \theta D
 - \kappa \ln \det (-\partial_t^2+r^2+D) + \kappa \ln \det(-\partial_t^2+r^2) \\
 &=& \frac{\mu}{2} D^2 - \theta D - \kappa \sqrt{r^2+D} + \kappa \, r
 . \label{LeffconstDx}
\end{eqnarray}
By definition, the function $U$ of (\ref{susyconstr:oneparticle})
equals minus the coefficient of the term linear in $D$, i.e.\
$U=-\partial_D L|_{D=0}$, so we indeed get $U(r)=\theta + \kappa /
2r$, with $r=|\sx|$. Because of the non-renormalization theorem,
this also determines the remainder of the first order Lagrangian
$L^{(1)}$. In particular, a magnetic interaction is generated
equivalent to the field of a charge $\kappa$ Dirac monopole. The
second order term can also be read off from (\ref{LeffconstDx}).
The ``metric'' or ``effective mass'' factor is
$\mu(r)=\partial_D^2 L |_{D=0} = \mu + \kappa/4 r^3$. Comparing
this to the general bosonic Lagrangian (\ref{secondorder}), we get
for the ``K\"ahler potential'' $K(r)=\mu r^2/3 - \kappa \ln r /
2r$, which fixes the second order term \cite{DE}. To quadratic
order in velocities and $D$, this gives the following low energy
effective bosonic Lagrangian:
\begin{equation} \label{LVeff}
 L_{B,eff} = \frac{1}{2} \bigl( \mu + \kappa/4 r^3 \bigr) (\dot{\sx}^2 +
 D^2) - (\theta + \kappa/2r) D - \kappa \sA^d \cdot \dot{\sx} \, ,
\end{equation}
where $\sA^d$ is once again the unit Dirac potential
(\ref{dirpot}). Here $\kappa = \langle S_1,S_2 \rangle$ and
$\theta = \mu (\alpha_2-\alpha_1)$. Within our small phase
approximation, this is exactly what we had in supergravity (see
e.g. (\ref{sugraid})), since $\sin(\Delta \alpha) \approx \Delta
\alpha$ when $\Delta \alpha \ll 1$. Comparing with the
supergravity probe Lagrangian (\ref{extsugraprobeL}), we see that
on the other hand the quadratic part of the Lagrangian, unlike the
linear part, does not match. This was to be expected, since
supersymmetry does not sufficiently constrain these terms in the
Lagrangian, so there is no reason to expect agreement between the
substringy and the supergravity regimes at this order.

\FIGURE[t]{\centerline{\epsfig{file=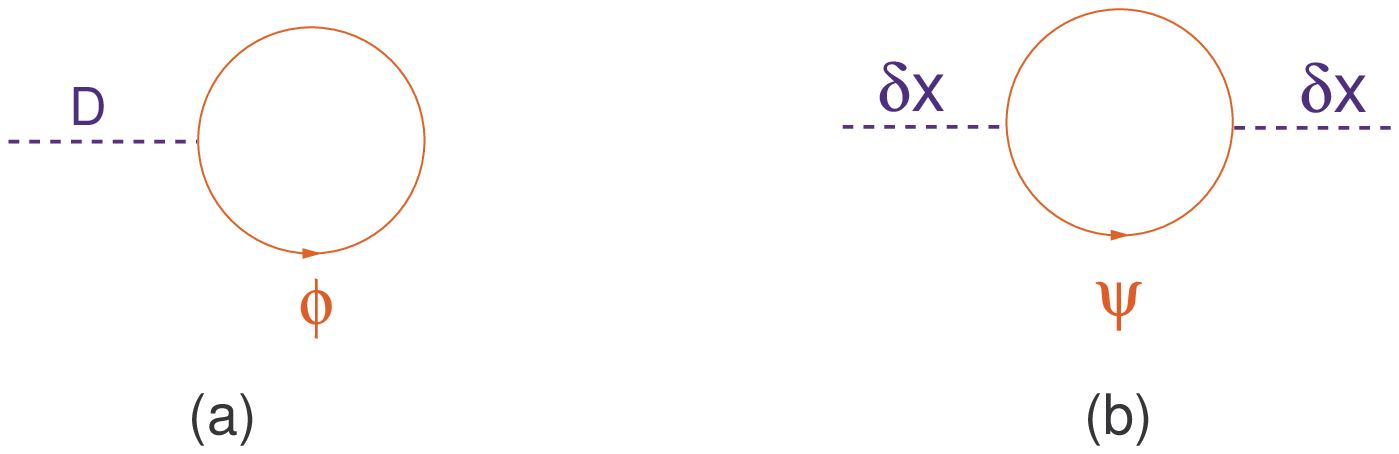,height=4.3cm}}
\caption{(a): the diagram generating $U$. (b): diagram generating
$\sA(\sx+\delta \sx) \cdot d \, \delta \sx$ to second order in
$\delta \sx$. }\label{diagrams}}

All this can be verified by computing one loop diagrams. For
example the correction to the FI term, or in other words the $1/r$
part of $U$, comes from the diagram shown in fig.\ \ref{diagrams}
(a): indeed,
\[ \int \frac{1}{\omega^2 + r^2} \, d\omega \, \sim 1/r \, . \]
The factor $\kappa$ in $U$ comes from summing over the $\phi^a$.

To conclude, we see that integrating out the charged chiral quiver
modes yields an effective potential on the Coulomb branch,
restoring supersymmetry at its minimum, and reproducing the
features of the BPS particle mechanics discussed in section
\ref{sec:particles}. Moreover, the first order part $L^{(1)}$,
which determines the equilibrium separation and the magnetic
interaction, is exactly the same in large scale and substringy
regimes.

We now address the question in which (substringy) regime the
chiral modes can indeed be integrated out to arrive at a good
description of the low energy physics. From the discussion in
section \ref{sec:example} (see for instance
(\ref{higgspotential})), we know that the on-shell classical mass
of $\psi$ and $\phi$ is $r$ and $\sqrt{r^2+\Delta \alpha}$
respectively, with $\Delta \alpha \equiv \alpha_2 - \alpha_1$.
Therefore if $\Delta \alpha<0$ and we only want to integrate out
massive modes, we should restrict to positions with $r \gg
\sqrt{-\Delta \alpha}$. In particular, this should be the case for
the equilibrium radius $R=\frac{\kappa}{2\mu|\Delta\alpha|}$,
giving the condition
\begin{equation} \label{particlegood}
 \mu \, |\Delta \alpha|^{3/2} \ll \kappa.
\end{equation}
On the other hand, if we want to stay in the substringy (``gauge
theory'') regime, where the quiver model gives a reliable
description of the low energy physics, we need $R \ll 1$, i.e.\
\begin{equation}
 \mu \, |\Delta \alpha| \gg \kappa \, .
\end{equation}
For sufficiently small $\Delta \alpha$, the two conditions are
satisfied simultaneously for a certain range of $\mu$, or
equivalently for a certain range of the string coupling constant
$g_s$, since $\mu = c/g_s$ with $c$ a function of the Calabi-Yau
moduli only; more precisely $c = \zeta \sqrt{v}/l_s$, with
$\zeta=|Z_1| |Z_2|/(|Z_1|+|Z_2|)$ and $v$ defined as under
(\ref{planckstringrelation}).

\FIGURE[t]{\centerline{\epsfig{file=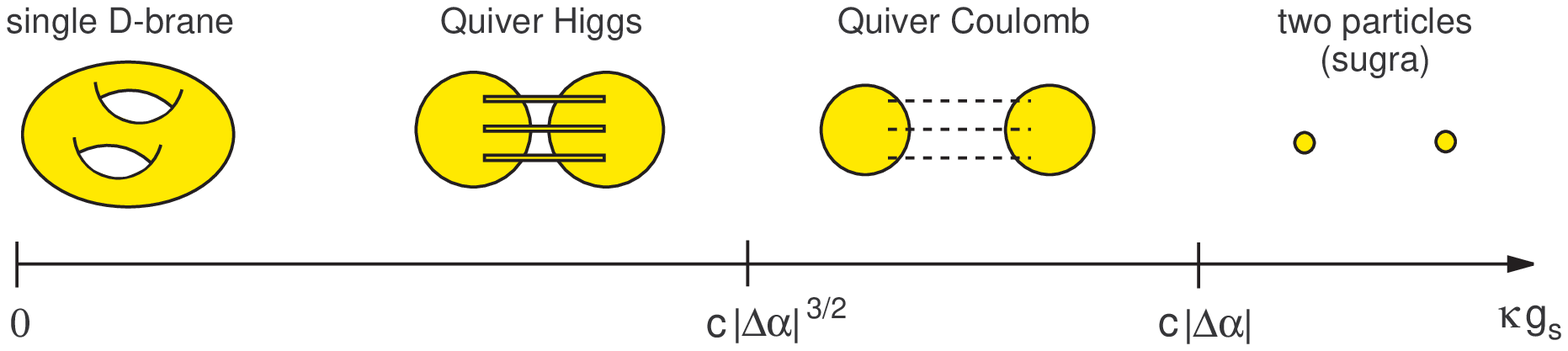,height=4cm}}
\caption{Metamorphosis of a single D-brane into a 2-particle
configuration. In the quiver quantum mechanical model, the wave
function lives on the Higgs branch when $\kappa g_s \ll c |\Delta
\alpha|^{3/2}$, and on the Coulomb branch (with effective
potential from integrating out the interconnecting strings) when
$\kappa g_s \gg c |\Delta \alpha|^{3/2}$. The substringy quantum
mechanical model is valid if $\kappa g_s \ll c |\Delta \alpha|$.
\label{transition}}}

Thus we see that by lowering the string coupling $g_s$ down to
zero, we first go smoothly and uneventfully from a macroscopic two
centered configuration in the large scale regime to a microscopic
two centered configurations in the substringy regime, but that if
we keep on lowering $g_s$, the strings stretched between the
branes in equilibrium become tachyonic and the two centered system
becomes (classically) unstable with respect to decay into a
configuration with nonzero $\phi$ --- in other words, we fall from
the Coulomb phase into the Higgs phase. Finally, at the extreme
classical limit $g_s=0$, the $\phi^a$ can be interpreted as the
moduli of a suitable geometric object, for instance a
supersymmetric D3-brane with $\kappa$ connecting throat moduli
$\phi^a$ \cite{slags}. The different regimes are illustrated in
fig.\ \ref{transition}.

Quantum mechanically, this phase transition is less sharp because
of quantum fluctuations. However, we can still identify a regime
in which the effective dynamics on the Coulomb branch captures the
low energy physics consistently and a complementary regime in
which the dynamics on the Higgs branch does so.

To have a consistent low energy description in terms of the vector
multiplets only, one needs the oscillator frequencies of the
integrated out chiral modes to be much higher than those of the
vectors (in the resulting effective dynamics). In the regime
(\ref{particlegood}), at the equilibrium distance, the frequency
of the chiral modes is approximately $\omega_c \sim
\frac{\kappa}{\mu |\Delta \alpha|}$, while from (\ref{LVeff})
(after integrating out D), it follows that the frequency of the
radial $\sx$-mode is $\omega_v \sim \frac{\mu |\Delta
\alpha|^2}{\kappa}$. So we have in this regime
\begin{equation}
 \frac{\omega_v}{\omega_c}
 \sim \frac{\mu^2 |\Delta \alpha|^3}{\kappa^2},
\end{equation}
and the condition for this to be small is precisely again
(\ref{particlegood}).

On the other hand, when $g_s \to 0$, one expects the more familiar
semi-classical description of the low energy dynamics of D-branes
to become valid, namely supersymmetric quantum mechanics on the
classical moduli space. To check this in the case at hand,
consider the Lagrangian after integrating out the auxiliary fields
from (\ref{relLV})-(\ref{relLC}):
\begin{eqnarray}
 L_{rel} &=& \frac{\mu}{2} ( {\dot{\sx}}^2
 + 2 i \bar{\lambda} \dot{\lambda}) + |\CD_t \phi^a|^2 + i \, \bar{\psi^a} \CD_t \psi^a \\
 &&  - |\sx|^2 |\phi^a|^2
 - \frac{1}{2 \mu} ( |\phi|^2 + \mu \Delta\alpha )^2
 \nonumber \\
 &&
 - \bar{\psi^a} \, \sx \cdot \ssigma \, \psi^a
 - i \sqrt{2} (\bar{\phi}^a \psi^a \epsilon \lambda -
 \bar{\lambda} \epsilon \bar{\psi}^a \phi^a) \, ,
\end{eqnarray}
As discussed earlier, when $\Delta \alpha < 0$, the classical
moduli space is non-empty. It is given by the zeros of the
potential $V$, i.e.\ $\sx=0$ and $|\phi|^2=-\mu \Delta \alpha$. If
we want the semi-classical picture of a particle moving in this
moduli space to be reliable, the spread of the particle wave
function out of the zero locus of $V$ has to be small compared to
the scale of the Mexican hat potential. More precisely, if we
write $\phi^a$ as $\phi^a = e^\sigma \phi_0^a$, with
$|\phi_0|^2=-\mu \Delta \alpha$, we should have $\langle \sigma^2
\rangle \ll 1$. Furthermore, to justify putting $\sx=0$ and thus
neglecting the $|\sx|^2|\phi^a|^2$ contribution to the potential
for $\phi$, we need $\langle \sx^2 \rangle \ll - \Delta \alpha$.

The semi-classical approximation consists of treating the $\sigma$
and $\sx$ modes as harmonic oscillators. For the $\sigma$-mode,
this is a harmonic oscillator with mass $\sim -\mu \Delta \alpha$
and spring constant $\partial_\sigma^2 V\, |_0 \sim \mu (\Delta
\alpha)^2$. Then $\langle \sigma^2 \rangle \sim \frac{1}{\mu
|\Delta \alpha|^{3/2}}$. So to have $\langle \sigma^2 \rangle \ll
1$, we need
\begin{equation} \label{higgsgood}
 \mu |\Delta \alpha|^{3/2} \gg 1 \, .
\end{equation}
For the $\sx$-mode, we get a harmonic oscillator with mass $\mu$
and spring constant $-\mu \Delta \alpha$, so $\langle \sx^2
\rangle \sim \frac{1}{\mu |\Delta \alpha|^{1/2}}$. The condition
for this to be much smaller than $-\Delta \alpha$ is again
(\ref{higgsgood}). Note that this condition is
complementary\footnote{At large $\kappa$, there can be overlap, in
which case there are two consistent semi-classical descriptions.
The full wave function will then presumably be a superposition of
the two.} to $(\ref{particlegood})$, and is indeed satisfied when
we take $g_s \to 0$ (i.e.\ $\mu \to \infty$) while keeping $\Delta
\alpha$ constant.

Finally, in this regime, the $\sx$ and $\sigma$ modes and their
superpartners have frequencies $\sim \sqrt{-\Delta\alpha}$ and
both should therefore be integrated out if we want an effective
Lagrangian for the moduli space zero modes alone. This leaves us
with a $d=1$, $\CN=4$ nonlinear sigma model on $\CP^{\kappa-1}$.

\subsection{Coulomb quiver quantum mechanics} \label{sec:coulquiv}

In the quiver Coulomb regime of fig.\ \ref{transition}, and near
equilibrium, the second term in the metric factor in (\ref{LVeff})
is only a small correction. So as far as ground state counting is
concerned (which is what we are ultimately interested in), we can
safely drop this correction, and the full Lagrangian becomes of
the simple form (\ref{LV}). After eliminating the auxiliary
variable $D$, we get
\begin{equation} \label{LVeffsimple2}
 L_{V,eff} = \frac{\mu}{2} \dot{\sx}^2  + i \, \mu \, \bar{\lambda}
 \dot{\lambda} - \frac{1}{2 \mu}(\theta + \kappa/2r)^2  - k \sA^d \cdot \dot{\sx}
 - \frac{k \, \sx}{2 r^3} \cdot \bar{\lambda} \ssigma \lambda \, .
\end{equation}
The (anti-) commutation relations can be read off from this
Lagrangian, and in particular we have $\{
\lambda_\alpha,\lambda_\beta \} = 0$,
$\{\bar{\lambda}^\alpha,\bar{\lambda}^\beta \}=0$,
$\{\lambda_\alpha,\bar{\lambda}^\beta \} = \mu^{-1}
\delta_\alpha^\beta$. The $\lambda$-operators can be represented
as $4 \times 4$ matrices, acting on a vector space generated by
$|0\rangle, \bar{\lambda}^\alpha |0\rangle, \bar{\lambda}^1
\bar{\lambda}^2 |0\rangle$, with $\lambda_\alpha|0\rangle=0$. The
supersymmetry generators are
\begin{eqnarray}
 Q_\alpha &=& \sigma^{i,\beta}_\alpha \lambda_\beta D_i - \lambda_\alpha U(r) \label{susyQ1} \\
 \bar{Q}^\alpha &=& - \bar{\lambda}^\beta \sigma^{i,\alpha}_\beta D_i - \bar{\lambda}^\alpha
 U(r) , \label{susyQ2}
\end{eqnarray}
where $D_i=\partial_i + i \kappa A^d_i$ and $U(r)=\kappa/2r +
\theta$. The supercharges satisfy the usual algebra
$\{Q_\alpha,Q_\beta\}=0$, $\{\bar{Q}^\alpha,\bar{Q}^\beta\}=0$,
$\{Q_\alpha,\bar{Q}^\beta\} = 2 \delta_\alpha^\beta H$, with
Hamiltonian $H$ given by
\begin{equation} \label{Hamiltonian}
 H = \frac{1}{2\mu} {D_i}^2 + \frac{1}{2 \mu}(\theta + \kappa/2r)^2
 + \frac{k \, \sx}{2 r^3} \cdot \bar{\lambda} \ssigma \lambda \, ,
\end{equation}
A general wave function is of the form
\begin{equation}
 F = \Phi(\sx) |0\rangle + \Psi_\alpha(\sx) \bar{\lambda}^\alpha |0\rangle
 + \tilde{\Phi}(\sx) \bar{\lambda}^1 \bar{\lambda}^2 |0\rangle \, .
\end{equation}
Supersymmetric ground states are given by wave functions
annihilated by all four supercharges
(\ref{susyQ1})-(\ref{susyQ2}). One easily deduces that this
requires $\Phi=0,\tilde{\Phi}=0$, while $\Psi$ must satisfy
\begin{equation}
  (\sigma^i D_i - U) \Psi = 0
\end{equation}
This equation can be solved by standard separation of variables.
As expected, when $\kappa$ and $\theta$ have the same sign, there
is no normalizable solution. If they have opposite sign, say
$\kappa>0$ and $\theta<0$, we get, in spherical coordinates
$(r,\vartheta,\varphi)$,
\begin{equation}
 \Psi_m \sim r^{\frac{\kappa}{2}-1} e^{\theta r} e^{i m \varphi}
  (1-\cos \vartheta)^{\frac{m-1}{2}}
 (1+\cos \vartheta)^{-\frac{m+1}{2}+\frac{\kappa}{2}} \,
 \left( \begin{array}{c} 1-\cos \vartheta \\ - e^{i \varphi} \sin \vartheta
 \end{array} \right) \, , \label{wavefunction}
\end{equation}
where $0 \leq m \leq \kappa-1$. These $\kappa$ solutions fill out
a spin $(\kappa-1)/2$ multiplet. Note that this is $1/2$ less than
what one would get from quantizing a spinless point particle in
the same setup. Physically, this is because the superparticle we
are considering here minimizes its energy by going into a spin
$1/2$ state aligned with the radial magnetic field, thus giving a
contribution of one spin quantum opposite to the intrinsic field
angular momentum. This explains also why the spin zero components
$\Phi$ and $\tilde{\Phi}$ vanish in a ground state.

To be complete, we also have to quantize the free center of mass
degrees of freedom, governed by the Lagrangian (\ref{comdof}). The
fermionic zeromodes give two spin zero singlets and one spin $1/2$
doublet. The total spin of the states and the supersymmetry
representations they carry are then obtained as the direct product
of the center of mass wave functions and the relative wave
functions. Thus the case $\kappa=1$ gives a (half) hypermultiplet,
$\kappa=2$ a vector multiplet, and so on.

Note that the radial probability density $p_r$ derived from the
wave functions (\ref{wavefunction}),
\begin{equation}
 p_r=\int |\Psi_m|^2 r^2 \sin
\vartheta \, d\vartheta d\varphi \,\, \sim r^{\kappa} e^{2 \theta
r}
\end{equation}
is peaked around the classical equilibrium point
$r=-\kappa/2\theta$, and that the width of the wave function
compared to the size of the configuration scales as
$1/\sqrt{\kappa}$, as one would expect.

\subsection{Higgs quiver quantum mechanics}
\label{sec:higgsquiver}

We now turn to the more familiar counting of supersymmetric ground
states in the Higgs regime, i.e. $\CN=4$ supersymmetric quantum
mechanics on the classical moduli space $\CM = \CP^{\kappa-1}$. We
denote the $\kappa-1$ bosonic degrees of freedom (complex
coordinates on $\CP^{\kappa-1}$) as $z^m$, and their superpartners
as $\chi^m$. As is well known, the quantum supersymmetric ground
states of a free particle moving in a K\"ahler manifold ${\cal M}$
are in one to one correspondence with the Dolbault cohomology
classes of $\CM$. This can be seen by identifying the quantized
fermionic operators with the following wedge and contraction
operations:
\begin{equation} \label{identifications}
 \chi^m_1 \to g^{m \bar{n}} \frac{\partial}{\partial (d\bar{z}^{\bar{n}})},
 \quad \chi^m_2 \to d z^m,
 \quad \bar{\chi}^{\bar{m},1} \to d\bar{z}^{\bar{n}},
 \quad \bar{\chi}^{\bar{m},2} \to g^{n \bar{m}} \frac{\partial}{\partial (dz^n)}
\end{equation}
For example $\chi^3_2 \cdot  dz^1 =  dz^3 \wedge dz^1$ and
$\bar{\chi}^3_2 \cdot  dz^1 \wedge dz^3 = g^{1 \bar{3}} dz^3 -
g^{3 \bar{3}} dz^1 $. With these identifications, the fermionic
canonical anticommutation relations are indeed satisfied: $\{
\chi^m_\alpha,\bar{\chi}^{\bar{n},\beta} \} = \delta_\alpha^\beta
\, g^{m\bar{n}}$, $\{\chi^m_\alpha,\chi^n_\beta \} = 0$, and
$\{\bar{\chi}^{\bar{m},\alpha},\bar{\chi}^{\bar{n},\beta} \} = 0$.
The supersymmetry operators are correspondingly identified as
follows:
\begin{equation}
 Q_1 \to \bar{\partial}^\dagger, \quad
 Q_2 \to \partial, \quad
 \bar{Q}^1 \to \bar{\partial}, \quad
 \bar{Q}^2 \to \partial^\dagger,
\end{equation}
so $\{ Q_\alpha,z^m \} = \chi_\alpha^m$ and so on. In this way
supersymmetric ground states are identified with the harmonic
representatives in the Dolbeault cohomology classes of the target
space.

The cohomology of $\CP^{\kappa-1}$ is well known; it has $\kappa$
elements: $1$, $\omega$, $\omega \wedge \omega$, $\ldots$,
$\omega^{\kappa-1}$, where $\omega$ is the K\"ahler form, $\omega
= -i g_{m\bar{n}} dz^m \wedge d\bar{z}^{\bar{n}}$. The
supersymmetric ground states we obtain in this way form a spin
$(\kappa-1)/2$ multiplet. Indeed, since the $z^m$ are invariant
under spatial $SO(3)$ rotations and the $\chi^m$ are obtained from
these by applying $Q$, which transforms in the {\bf 2}, the
spinors $\chi^m$ transform likewise in the ${\bf 2}$ under spatial
$SO(3)$ rotations. The corresponding angular momentum operator is
\begin{equation}
 {\bf S} = \frac{1}{2} \, g_{m\bar{n}} \, \bar{\chi}^{\bar{n}} \ssigma \chi^m.
\end{equation}
With the identifications (\ref{identifications}) and using the
anticommutation relations, this gives:
\begin{eqnarray}
 S^3 &=& \frac{1}{2} \bigl( d\bar{z}^{\bar{n}}
 \frac{\partial}{\partial \, d\bar{z}^{\bar{n}}}
  + d z^m \frac{\partial}{\partial \, dz^n} \bigr) -
  \frac{\kappa-1}{2} \\
 &=& \frac{1}{2} ( \mbox{form number} -
  \mbox{dim}_{\IC}) \label{dimspinrel} \\
 S^+ &=& - g_{m\bar{n}} dz^m \wedge d\bar{z}^{\bar{n}} = -i \omega \\
 S^- &=& g^{m\bar{n}} \frac{\partial}{\partial \, dz^m} \wedge
 \frac{\partial}{\partial \, d\bar{z}^{\bar{n}}} \,.
\end{eqnarray}
An $SU(2)$ action of this form always exists on the cohomology of
K\"ahler manifolds; it is called the Lefschetz $SU(2)$. In this
case we see that the Lefschetz $SU(2)$ coincides with the spatial
$SU(2)$, and the cohomology of $\CP^{\kappa-1}$ forms a spin
$(\kappa-1)/2$ multiplet under this group action, as announced.

So we get the same supersymmetric ground state degeneracy and spin
as in the corresponding quantum mechanics on the Coulomb branch.
By sending $g_s$ to zero, the ground states simply underwent a
continuous metamorphosis from living on the Coulomb branch to
living on the Higgs branch, or in other words from a spatial
molecular form to a geometric D-brane form.

\subsection{General case}

The generalization to arbitrary quivers is in principle
straightforward. Let us take all constituent branes separated so
stretched strings become massive. Integrating out strings between
copies of the same brane will not produce an interaction at first
velocity order because fermionic and bosonic string modes have the
same mass. A string stretched between different species connected
by arrow in the corresponding quiver diagram on the other hand
generically does produce an effective first order interaction,
with sign determined by the orientation of the corresponding arrow
in the quiver. This can also be verified directly from the
abelianized quiver Lagrangian at the end of appendix
\ref{app:genquiv}. For each such string we integrate out, a two
body interaction just like the one we had for the two particle
case is induced. Thus the resulting contribution to the effective
first order interaction Lagrangian for the position multiplets is
simply a sum over all pairs of particles, with for each pair the
two body interaction we found previously in the toy example. The
terms in the sum are furthermore weighted by the intersection
product $\langle v,w \rangle$ of the corresponding objects of type
$v$ and $w$, which is equal to the number of arrows $v \to w$
minus number of arrows $w \to v$. So we reproduce the Lagrangian
(\ref{multiL11})-(\ref{multiL12}) with $\kappa_{wv}=\langle v,w
\rangle$.

%To illustrate this procedure, consider the quiver of fig.\
%\ref{quivnoloops}.

One could worry about the possible presence of a superpotential
(if the quiver has closed loops). If $W \neq 0$, the Lagrangian is
no longer quadratic in the chiral multiplets, and integrating them
out will involve higher loop diagrams. However, because of the
nonrenormalization theorem for the first order part of the
Lagrangian, we are allowed to compute this term at arbitrary large
separations, where the $|\Delta x|^2 |\phi|^2$ mass terms are
certainly dominant and the superpotential terms can be ignored.
Hence the superpotential will not affect $L_{eff}^{(1)}$.

As for the two center case, sending $g_s$ all the way down to zero
while keeping all other parameters fixed makes the multicentered
structure collapse to mutual distances where bosonic strings
become tachyonic and the description in terms of separate centers
breaks down. Again, the appropriate picture then becomes the Higgs
branch of the quiver quantum mechanics.\footnote{Note that there
can be intermediate stages which look like multicentered
configurations with as centers bound states of branes in their
Higgs phases.}

However, in the opposite direction, starting at $g_s=0$ with a
classical wrapped D-brane bound state localized at a single point
in space, and increasing the string coupling constant, things are
not that simple. In general, the state will not necessarily ``open
up'' and become a multicentered configuration of partons. Instead,
it could for example stay centered in one point and turn into a
black hole, or it could split into two black holes at a certain
equilibrium distance, or into a black hole surrounded by a cloud
of partons, and so on. In other words, there are many more D-brane
states than there are multi-particle states (of the kind we have
been describing), and this discrepancy is shows up in the form of
emerging horizons.

\FIGURE[t]{\centerline{\epsfig{file=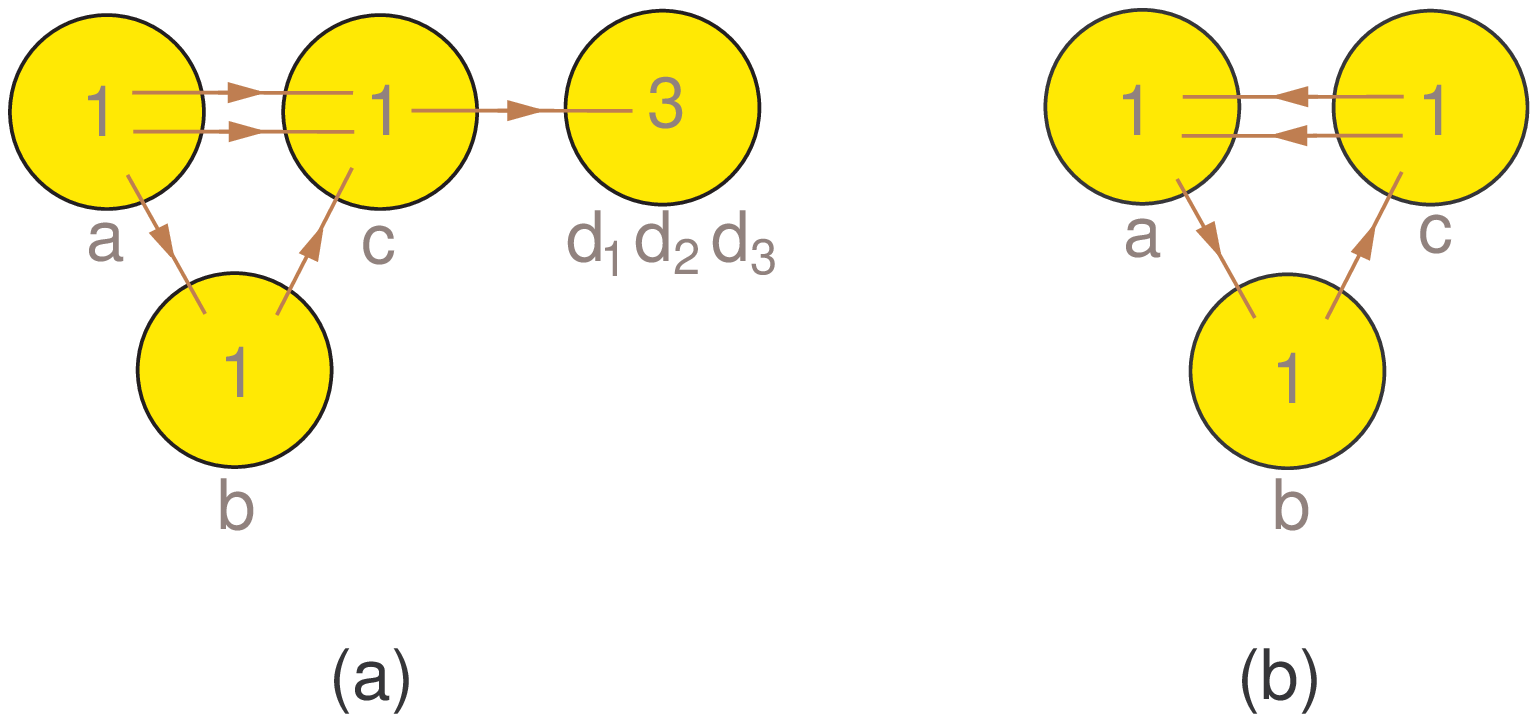,height=5cm}}
\caption{(a): A quiver with no loops and dimension vector
(1,1,1,3). The arrows define the ordering $a<b<c<d$. (b): A quiver
with a loop. \label{quivnoloops}}}

The obvious question is then: what distinguishes between the
different possibilities? We don't know the complete answer to that
question, but we can argue for a certain class of quivers that
they will certainly open up into a multicentered state when $g_s$
increases, namely for quivers without closed loops, such as the
one shown in fig.\ \ref{quivnoloops}a. This excludes quiver
diagrams with arrows going from a node to itself, diagrams with
arrows going in two directions between a pair of nodes (fig.\
\ref{quivers}b) and diagrams such as fig.\ \ref{quivnoloops}b or
\ref{quivers}c. From (\ref{Dflatness}) and (\ref{multieq}), it is
not hard to see that on both the D-brane and the particle side,
having no closed loops implies that the moduli space is compact if
the $\theta_v$ are finite, in the sense that the $\phi^a$ are
bounded from above and the separations between mutually nonlocal
particle species bounded from below\footnote{and from above for
coprime dimension vector and sufficiently generic moduli such that
none of the constituents has its phase equal to the overall phase
of the system}, respectively. This is because having no closed
loops implies that we can consistently define a partial ordering
of the nodes according to the direction of the arrows. A would-be
divergence in the left hand side of (\ref{Dflatness}) or
(\ref{multieq}) at a certain node would propagate through the
chain of equations to a maximal or a minimal node, where a
divergence is manifestly impossible since the left hand sides of
(\ref{Dflatness}) and (\ref{multieq}) have only positive or only
negative contributions there. As an example, consider the
equilibrium equations (\ref{multieq}) for the quiver of fig.\
\ref{quivnoloops}, which has the ordering $a<b<c<d$:
\begin{eqnarray}
 \frac{1}{r_{ab}} + \frac{2}{r_{ac}} &=& 2 \theta_a \\
 \frac{1}{r_{bc}} - \frac{1}{r_{ab}} &=& 2 \theta_b \\
 \sum_{j=1}^3 \frac{1}{r_{a,dj}} - \frac{2}{r_{ac}} - \frac{1}{r_{bc}} &=& 2 \theta_c \\
 -\sum_{j=1}^3 \frac{1}{r_{a,dj}} &=& 2 \theta_d.
\end{eqnarray}
It is clear that the first and the last equation prevent the full
set of $r_{pq}=|\sx_p-\sx_q|$ to become arbitrary small. Thus, if
two particles mutually interact, they stay away from each other.

Also, having no arrows from a node to itself means that the
constituents have no internal moduli of their own. Since no moduli
means no entropy, one therefore expects that the corresponding
particles, when put together, do not form black holes. This is
confirmed by examples \cite{branessugra,DGR}, where they form,
through the enhan\c{c}on mechanism, ``empty'' holes instead
(giving interaction potentials like fig.\ \ref{potential}b and
\ref{potential}c). Black holes can form if mutually interacting
particles of different species come infinitely close in coordinate
distance, but as we just saw this is not possible if the quiver
has no closed loops. So we can apply the reasoning we made for the
two-particle case, without black hole or non-compactness
complications, and the transition of states living on the Higgs
branch to states living on the Coulomb branch when $g_s$ increases
should be equally clean.

This is in contrast to cases with loops, such as fig.\
\ref{quivnoloops}b. The equilibrium equations for this example are
\begin{eqnarray}
  \frac{1}{r_{ab}} - \frac{2}{r_{ac}} &=& 2 \theta_a \\
 \frac{1}{r_{bc}} - \frac{1}{r_{ab}} &=& 2 \theta_b \\
  \frac{2}{r_{ac}} - \frac{1}{r_{bc}} &=& 2 \theta_c.
\end{eqnarray}
These equations always have solutions for arbitrarily small
$r_{pq}$, so the interacting particles can come arbitrarily close
to each other and form a black hole or at least a microscopic sort
of bound state that falls outside the type of states we have been
studying thus far. Note also that these configurations are more
stable than the cases without closed loops: there will always be
solutions to the above equations, no matter what values the
$\theta_v$ have.

Finally, on the D-brane side, closed loops and the potential
presence of a superporential go hand in hand. This suggests that
black hole formation and the appearance of superpotentials are
perhaps intrinsically linked, but we will not get into this in the
present work. Instead, we will focus on quivers without closed
loops, where the Coulomb-Higgs relation is most transparent.

Can we expect in general a detailed match between the
supersymmetric ground state degeneracies in both pictures, as we
found for the case of the quiver with two nodes, $\kappa$ arrows
and dimension vector $(1,1)$? As we will see below, this detailed
correspondence also holds for a number of more involved examples,
and it is tempting to conjecture that this is indeed the case in
general, but we have no proof of this. The problem is as usual
that during the continuous interpolation between the two regimes,
boson-fermion pairs of states could in principle be moved in or
out the ground state set. We will be able to show that in the
Higgs regime, the supersymmetric ground states are either all
fermionic or all bosonic. Showing the same for the Coulomb regime
would therefore establish the detailed correspondence, but how to
do this in general is an open question.

A quantity that should match in any case is of course the Witten
index, defined here as the number of bosonic minus the number of
fermionic ground states of the system before tensoring with the
trivial center of mass half-hypermultiplet (after tensoring the
index is trivially zero).

\subsection{A nontrivial test: the Hall halo} \label{sec:hallhalotest}

For our basic example, corresponding to a quiver with two nodes,
$\kappa$ arrows, and dimension vector $(1,1)$, we were able to
compute ground state degeneracies on both sides, with identical
results. With a bit more effort, the same can be done for
dimension vector $(1,N)$, providing a nontrivial test of the
proposed correspondence. On the Coulomb side, this system can be
thought of as a charge $\kappa$ magnetic monopole surrounded by
$N$ mutually non-interacting electrons of charge $1$, or in other
words a Hall halo. To construct wave functions for this system,
one has to make antisymmetric combinations of the one-particle
wave function (\ref{wavefunction}). A very similar problem was
considered and explicitly solved in \cite{qhall}, with the purpose
of studying the quantum Hall effect on a sphere. The only
difference is that the one-particle wave function of \cite{qhall}
is that of a spinless particle, while ours is that of a
superparticle frozen in a preferred energy-minimizing spin $1/2$
state (as discussed in section \ref{sec:coulquiv}), thus lowering
the maximal spin with $1/2$ and the ground state degeneracy with 1
compared to the spinless particle case. This can be taken into
account effectively by simply subtracting one unit from the number
of flux quanta wherever this quantity appears in \cite{qhall}.

In particular, this gives us without further effort a generating
function for the number of ground states with given spin $J_3$
along the $3$-axis.\footnote{Here and in what follows, we count
the states before tensoring with the trivial center of mass
half-hypermultiplet.} If we denote by $n_L$ the number of
supersymmetric ground states with $J_3 = L/2 - N(\kappa-N)/2$,
this is \cite{qhall}:
\begin{equation} \label{genfunct}
 G(t) \equiv \sum_L n_L \, t^L = \frac{\prod_{j=1}^\kappa (1-t^{2j})}
 {\prod_{j=1}^N (1-t^{2j}) \prod_{j=1}^{\kappa-N} (1-t^{2j})} \, .
\end{equation}
The number of spin $j=L/2 - N(\kappa-N)/2$ multiplets equals $n_L
- n_{L-1}$, and the total degeneracy is $n_{tot}=G(1)={\kappa
\choose N}$. The latter can be easily understood from the
exclusion principle: since our $N$ electrons are mutually
noninteracting fermions, and the one-particle degeneracy is
$\kappa$, states are labeled by filling up $N$ slots out of a
total of $\kappa$. Also because of the exclusion principle, the
maximal spin ($N(\kappa-N)/2$) is lower than the naive
supergravity expectation given by (\ref{sugraspin}) ($\kappa
N/2$). The highest spin in supergravity is obtained by putting all
electrons on top of each other, which is not allowed quantum
mechanically.

As an example, the particle system corresponding to the $\kappa=5$
quiver of fig.\ \ref{quivers}a, with dimension vector $(1,N)$, has
a single spin $j=0$ ground state for $N=0$, a spin $2$ multiplet
for $N=1$, and a spin $3$ multiplet plus a spin $1$ multiplet for
$N=2$. The cases $N=3,4,5$ have the same structure as $N=2,1,0$
respectively. There are no quantum supersymmetric ground states
for $N>5$, which is again a direct consequence of the exclusion
principle. On the other hand, as shown in \cite{DGR}, there are
\emph{classical} supersymmetric ground states with $N>5$ for this
system (embedded in supergravity). It was also pointed out there
that these classical BPS configurations are not stable under
monodromies, despite the fact that there is no line of marginal
stability crossed. One of the proposed ways out of this paradox
(also known as the ``s-rule problem'') was that there was simply
no BPS state with $N>5$ at the \emph{quantum} level to begin with.
From our analysis here, we see that this was indeed the correct
interpretation. Closely related problems were encountered for
example in \cite{argyres}. The relation between the s-rule, the
exclusion principle and geometry was also pointed out (in a
different context) in \cite{BG}.

Finally note that the number of ground states grows exponentially
fast with $\kappa$ at constant filling fraction $\nu=N/\kappa$,
since $n_{tot}={\kappa \choose N} \approx \exp[\kappa(\nu \ln
\nu^{-1} + (1-\nu)\ln(1-\nu)^{-1})]$.

We now turn to the Higgs branch description and see if the ground
states there match those of the Coulomb regime. As argued under
(\ref{Dflatex1})-(\ref{Dflatex2}), the moduli space of the quiver
with two nodes, $\kappa$ arrows and dimension vector $(1,N)$ is
the Grassmannian $Gr(N,\kappa)$, the space of all $N$-planes in
$\IC^\kappa$. The cohomology of this space is classically known.
The generating function for the betti numbers $b_L$, also known as
the Poincar\'e polynomial, is \cite{bott}:
\begin{equation} \label{grassmanpp}
  P(t) \equiv \sum_L b_L \, t^L = \frac{\prod_{j=1}^\kappa (1-t^{2j})}
 {\prod_{j=1}^N (1-t^{2j}) \prod_{j=1}^{\kappa-N} (1-t^{2j})} \, .
\end{equation}
Furthermore, because of (\ref{dimspinrel}) and $\dim
Gr(N,\kappa)=N(\kappa-N)$, we have that an $L$-form has spin $J_3
= L/2 - N(\kappa-N)/2$. Comparing this to (\ref{genfunct}) and the
spin assignment there, we find perfect agreement between the two
different pictures, even though the counting problems look very
different at first sight. The framework presented in this paper
provides a conceptual understanding of this remarkable match.

%%%%%%%%%%%%%%%%%%%%%%%%%%%%%%%%%%%%%%%%%%%%%%%%%%%%%%%%%%%%%%%%%%%%%%%%
\section{Further tests and applications}\label{sec:LLL}
\setcounter{equation}{0}
%%%%%%%%%%%%%%%%%%%%%%%%%%%%%%%%%%%%%%%%%%%%%%%%%%%%%%%%%%%%%%%%%%%%%%%%

\subsection{Counting cohomology classes of arbitrary quiver varieties}

Thanks to a recent result of Reineke \cite{reineke}, it has become
possible, at least in principle, to compute explicitly the Betti
numbers of (semi-)stable quiver moduli spaces for arbitrary
quivers $Q=(V,A)$ without closed loops, for arbitrary dimension
vectors $\sN = (N_v)_{v \in V}$, at arbitrary values of the FI
parameters $\theta_v$. The proof delves deep into the mathematics
of finite fields, quantum groups and the Weil conjectures, but
fortunately the final result can be stated as a down-to-earth
explicit formula for the Poincar\'e polynomial $P(t)=\sum_L b_L \,
t^L$, i.e.\ the generating function of the Betti numbers $b_L =
\dim H^L\left( {\cal M}_d^{ss}(Q,\sN,\theta),\IC \right)$.
Denoting $[N,t] \equiv \frac{t^{2N}-1}{t^2-1}$ and $[N,t]! \equiv
[1,t][2,t]\ldots[N,t]$, we have \cite{reineke}
\begin{equation} \label{thebigformula}
 P(t)=(t^2-1)^{1-\sum_v N_v} \, t^{-\sum_v N_v(N_v-1)} \,
 \sum_{\sN^*}(-1)^{s-1} t^{2\sum_{k\leq l}\sum_{v \rightarrow
w}N^l_v N^k_w} \prod_{k,v} ([N^k_v,t]!)^{-1}
\end{equation}
where the sum runs over all ordered partitions of $\sN$ by
non-zero dimension vectors $\sN^*=(\sN^1\ldots \sN^s)$ (i.e.\
$\sN=\sum_{k=1}^s \sN^k$, $\sN^k \neq 0$), satisfying
$\theta(\sum_{l=1}^k \sN^l) > 0$ for $\kappa=1\ldots s-1$. Note
that only even powers of $t$ occur. Physically this means that the
supersymmetric ground states are either all bosonic or all
fermionic (as usual before tensoring with the center-of-mass
multiplet), and in particular that the Witten index (i.e.\ the
Euler characteristic of the moduli space) equals the total number
of ground states.

\subsection{Bound states of $N_1$ ``monopoles'' with $N_2$
``electrons''} \label{sec:ab}

The quiver with two nodes and $\kappa$ arrows from the first to
the second node, and dimension vector $\sN=(N_1,N_2)$ corresponds
to a system of $N_1$ particles of one type and $N_2$ particles of
another type, with intersection product between the two kinds of
particles equal to $\kappa$ --- for example a system of $N_1$
``monopoles'' of charge $\kappa$ and $N_2$ ``electrons'' of charge
$1$. These particles are bound together (in the stable case
$\theta_2<0$) by a potential of the form $V \sim (1/r + 2
\theta_2/\kappa)^2$ between monopoles and electrons. We take $N_1$
and $N_2$ coprime, so stability and semi-stability are equivalent.
(Physically, a common divisor $d$ means that there is no potential
preventing the state to split into $d$ pieces of charge $\sN/d$,
giving rise to potential threshold bound state subtleties, which
we wish to avoid in this paper.)

In this case the formula (\ref{thebigformula}) can be simplified
to \cite{reineke}:
\begin{eqnarray} \label{bigformulabis}
 P^{\kappa}_{N_1,N_2}(t) &=&
 (t^2-1)^{1-N_1}\, \, t^{-N_1(N_1-1)}
 \nonumber
 \\
 && \times \sum_{N_1^*,N_2^*}(-1)^{s-1} t^{2\sum_{k<l}(\kappa N_1^l-N_2^l) N_2^k}
 \prod_{k=1}^s
 \frac{[\kappa N_1^k,t]!}{[N_1^k,t]![N_2^k,t]![\kappa N_1^k-N_2^k,t]!}
\end{eqnarray}
where the sum runs over all partitions $N_1^*=(N_1^1\ldots
N_1^s)$, $N_2^*=(N_2^1\ldots N_2^s)$ of $N_1$ resp.\ $N_2$ such
that $N_1^k \neq 0$ for all $k$, and
$(N_1^1+\ldots+N_1^k)/N_1>(N_2^1+\ldots+N_2^k)/N_2$ for all
$k=1\ldots s-1$.

As a check, consider the Hall halo case $N_1=1$. Then the formula
collapses simply to
\begin{equation}
 P^\kappa_{1,N_2}(t)= \frac{[\kappa,t]!}{[N_2,t]![\kappa-N_2,t]!},
\end{equation}
reproducing (\ref{grassmanpp}).

The formula (\ref{bigformulabis}) for $N_2 \geq 2$ gives new
predictions for the structure of monopole-electron BPS bound
states. This is a pretty nontrivial result, as even the
\emph{classical} ground state configurations are hard to obtain
explicitly. To be concrete, let us consider the example
$N_1=2,N_2=3$ with $\kappa=5$. This example appears in the context
of type IIA string theory compactified on the Quintic, see fig.\
\ref{quivers}. There are three contributing partition pairs
$(N_1^*;N_2^*)$, namely $((2);(3))$, $((1,1);(1,2))$ and
$((1,1);(0,3))$. The resulting Poincar\'e polynomial is, with $x
\equiv t^2$:
\begin{eqnarray*}
 P^5_{2,3} &=& 1 + x + 3\, x^2 + 4\, x^3 + 7\, x^4 + 9\, x^5 + 14\, x^6 + 16\, x^7 +
    20\, x^8 + 20\, x^9 \\
    && + 20\, x^{10} + 16\, x^{11} + 14\, x^{12} + 9\, x^{13} +
    7\, x^{14} + 4\, x^{15} + 3\, x^{16} + x^{17} + x^{18} .
\end{eqnarray*}
The organization of ground states in spin multiplets can be read
off directly from this polynomial, and is given in the following
table, where $j$ is the spin quantum number, $n_j$ the number of
spin $j$ multiplets in the ground state Hilbert space, and $d_j =
n_j (2j+1)$ the total dimension the spin $j$ multiplets occupy.
The total number of states is $170$.

\begin{center}
 \begin{tabular}{|r||l|l|l|l|l|l|l|l|l|l|}
 \hline
 $j$   & 9 & 8 & 7 & 6 & 5 & 4 & 3 & 2 & 1 & 0 \\
 \hline
 \hline
 $n_j$ & 1 & 0 & 2 & 1 & 3 & 2 & 5 & 2 & 4 & 0 \\
 \hline
 $d_j$ & 19 & 0 & 30 & 13 & 33 & 18 & 35 & 10 & 12 & 0 \\
 \hline
\end{tabular}
\end{center}

The full supermultiplet thus generated is obtained by taking the
direct product with the center of mass wave function, which
consists of two spin 0 singlets and one spin $1/2$ doublet. The
maximal total spin is thus $9+1/2$.

\subsection{BPS states in $\CN=2$ $SU(2)$ Yang-Mills}

A wide variety of constructions of BPS states in $\CN=2$ super
Yang-Mills theories exist, including the two pictures considered
in this paper. For the multi-particle picture in the (abelian) low
energy effective field theory, see for instance \cite{argyres}.
For the quiver D-brane picture, see in particular \cite{fiol},
where the full set of (classical) BPS states was built from a
certain set of parton branes, linked together by quiver diagrams
derived from the orbifold construction associated to this theory
through geometric engineering. In this construction, the number of
arrows between two nodes again equals the DSZ intersection product
of the objects corresponding to the nodes.

We will first consider the pure $SU(2)$ case. The low energy
effective theory, solved exactly by Seiberg and Witten in
\cite{SW}, is a $U(1)$ gauge theory. The theory has a one complex
dimensional moduli space, parametrized by a holomorphic coordiante
$u$. This space has three singularities: one at infinity, one at
$u=1$, and one at $u=-1$. In the quiver picture \cite{fiol}, there
are two partons, corresponding to a monopole and a dyon, which
become massless at $u=1$ and $u=-1$ respectively. The intersection
product of the two partons is $2$, giving a quiver with two nodes
and two arrows between them, a special case of the class of
quivers studied in section \ref{sec:ab}. The entire classical BPS
spectrum of the theory can be built as stable representations of
this quiver with various dimension vectors. As we saw, the
stability depends on the moduli through the FI coefficient
$\theta$. In this theory $\theta=0$ on an ellipse-like line
containing the two singularities, $\theta<0$ outside the line, and
$\theta>0$ inside. So to have a stable representations, we need to
be outside this line of marginal stability, i.e. in the weak
coupling region. Still, only a limited number of dimension vectors
$(n,m)$ support stable representations there. This follows already
from simple counting of the degrees of freedom modulo complexified
gauge transformations, giving for the expected dimension of the
quiver moduli space $d(n,m)=2 n m - n^2 - m^2 + 1=1-(n-m)^2$,
which is zero for $n-m=\pm 1$ and one for $n=m$.

\FIGURE[t]{\centerline{\epsfig{file=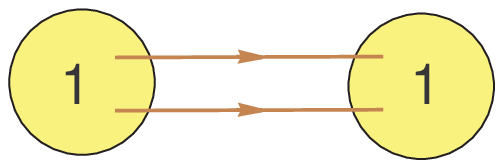,height=2.6cm}}
\caption{Quiver representation of a $W^\pm$ boson in $\CN=2$
$SU(2)$ Yang-Mills} \label{su2vect}}

The case $n=m=1$ is the only one with a nontrivial moduli space
and $\mbox{g.c.d.}(n,m)=1$. It corresponds to the $W^{\pm}$ gauge
boson. The corresponding quiver representation is shown in fig.\
\ref{su2vect}. The moduli space is $\CP^1$, so we get a spin $1/2$
multiplet of BPS ground states, which multiplied with the center
of mass states gives an $\CN=2$ vector multiplet, in agreement
with the interpretation of this particle as a gauge boson. The
quiver representations corresponding to the tower of dyons have a
zero dimensional moduli space, so they all come in
hypermultiplets, in agreement with their interpretation as cousins
of the parton monopole and dyon.

The particle picture (in the abelian low energy effective field
theory) should give the same result if the proposed duality is
correct. Unfortunately, even finding the classical moduli space in
this picture is very hard if more than one of each of the partons
is present, and we will not attempt to solve this problem here.
The $(1,n)$ or $(n,1)$ cases on the other hand are again the Hall
halo cases, and we found indeed that those agree with the quiver
moduli picture. It would be interesting to extend this analysis to
the general $(n,m)$ case in the particle picture, and in
particular to show how quantum mechanics corrects the fact that
one finds way too many BPS bound states classically in this
approach \cite{argyres}. As was the case for the Hall halo, this
is presumably due to the exclusion principle, but we do not know
how exactly this comes about.

\subsection{The Stern-Yi dyon chain}

In \cite{sy}, Stern and Yi studied the ground state counting
problem for a collection of $k+1 \leq N$ distinct dyons in $\CN=2$
$SU(n)$ Yang-Mills with magnetic charges $m_v$ given by an
irreducible (sub)set of simple roots,
$\beta_1,\ldots,\beta_{k+1}$. We take the roots normalized to
satisfy the relations $\beta_v^2=2$, $\beta_v \cdot \beta_{v+1} =
-1$, and $\beta_v \cdot \beta_w = 0$ for $|v-w|>1$, with the dot
denoting the inner product on weight space. The electric charges
of the dyons are $e_v=n_v \beta_v$. The DSZ intersection product
of two (magnetic, electric) charges $(m,e)$ and $(m',e')$ is
$\langle (m,e),(m',e') \rangle \equiv m \cdot e' - e \cdot m'$.
Thus we have for the above charges that $\langle
(m_v,e_v),(m_{v+1},e_{v+1})\rangle=n_{v}-n_{v+1} \equiv \kappa_v$,
and all other intersection products zero. Thus we obtain a
``chain'' of dyons, which can be represented by a quiver with
$k+1$ nodes, $\kappa_v$ arrows from node $v$ to node $v+1$ if
$\kappa_v>0$, $-\kappa_v$ arrows from node $v+1$ to node $v$ if
$\kappa_v<0$, and dimension vector $(1,1,\ldots,1)$, as shown in
fig.\ \ref{chain} for $k=4$.

The Witten index ${\cal I}$ for this chain was computed in
\cite{sy} in the framework of the moduli space approximation to
the low energy dynamics of dyons as solitons in the classical
nonabelian Yang-Mills theory. The main result was that, depending
on the moduli, either ${\cal I}=0$, or ${\cal I} = \prod_{v=1}^k
|\kappa_v|$. It was also conjectured that four times this index is
actually equal to the total number of BPS states (the four coming
from the quantization of the center of mass fermions).

\FIGURE[t]{\centerline{\epsfig{file=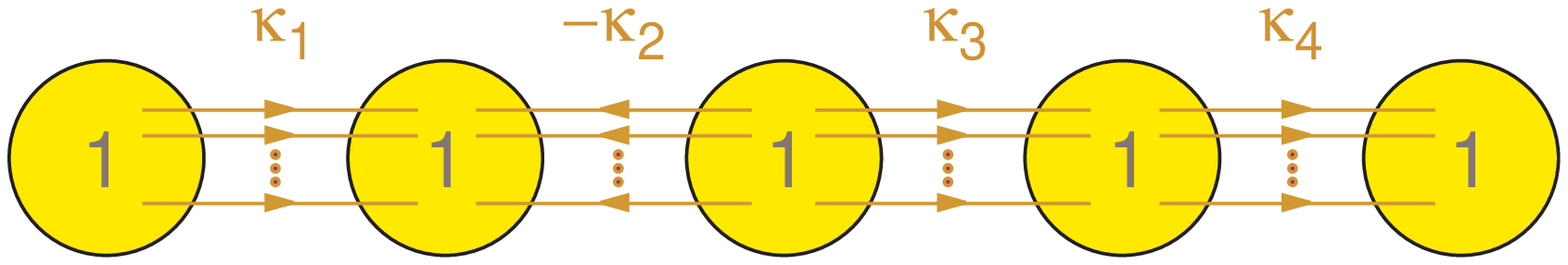,height=2.6cm}}
\caption{Quiver diagram corresponding to the $k=4$ Stern-Yi dyon
chain with $\kappa_1,\kappa_3,\kappa_4>0$ and $\kappa_2<0$.
\label{chain}}}

This result is reproduced in both of our pictures. We start with
the quiver moduli space picture. An arrow $a$ between $v$ and
$v+1$ corresponds to a complex variable $\phi^a_{v,v+1}$, where
$a=1,\ldots,|\kappa_v|$. The moduli space $\CM$ is, using the
D-flatness description (\ref{Dflatness}):
\begin{equation}
 \CM=\left\{ \phi: s_v \sum_{a=1}^{|\kappa_v|} |\phi^a_{v,v+1}|^2 - s_{v-1} \sum_{a=1}^{|\kappa_{v-1}|}
 |\phi^a_{v-1,v}|^2 = \theta_v \quad \mbox{for all nodes }
 v=1,\ldots,k+1 \right\} \, / \, G,
\end{equation}
where $\kappa_0 \equiv 0$, $\kappa_{k+1} \equiv 0$,
$s_v=\mbox{sign } \kappa_v$, the gauge group $G=U(1)^{k+1}$, and
$\theta_v = m_v (\alpha_v-\alpha_0)$ as in (\ref{FIgen}). This is
equivalent to the equations
\begin{eqnarray*}
 s_k \sum_{a=1}^{|\kappa_k|} |\phi^a_{k,k+1}|^2 &=& - \theta_{k+1} \\
 s_{k-1} \sum_{a=1}^{|\kappa_{k-1}|} |\phi^a_{k-1,k}|^2 &=& - (\theta_k + \theta_{k+1}) \\
 &\cdots& \\
 s_1 \sum_{a=1}^{|\kappa_1|} |\phi^a_{1,2}|^2 &=& - (\theta_2 + \theta_3 +
 \cdots + \theta_{k+1}) = \theta_1
\end{eqnarray*}
modulo $G$. A solution only exists if the partial $\theta$-sums
$s_{l-1} \sum_{v={l}}^{k+1} \theta_v$, $l \geq 2$ are all
negative. (As before, we discard nongeneric cases, i.e.\ having
some of the partial sums equal to zero, to avoid complications of
singularities and threshold bound states.) This is of course just
the $\theta$-stability condition discussed in section
\ref{quivmath}.\footnote{It should also match the stability
condition of \cite{sy}, but we didn't check this.} If this is
satisfied, the moduli space is
\begin{equation}
 \CM = \CP^{|\kappa_1|} \times \CP^{|\kappa_2|} \times \cdots \times \CP^{|\kappa_k|}.
\end{equation}
The Poincar\'e polynomial of $\CM$ is therefore
\begin{equation}
 P(t) = \prod_{v=1}^k \frac{t^{2 |\kappa_v|}-1}{t^2-1} \, ,
\end{equation}
from which we can directly read off the spin multiplet structure.
In particular this gives for the total number of BPS ground states
\begin{equation}
 \dim {\cal H}_{BPS} = 4 \times \prod_{v=1}^k |\kappa_v| \, ,
\end{equation}
where the factor of four comes again from the center of mass
degrees of freedom. This is in exact agreement with \cite{sy}, and
confirms their conjecture.

\FIGURE[t]{\centerline{\epsfig{file=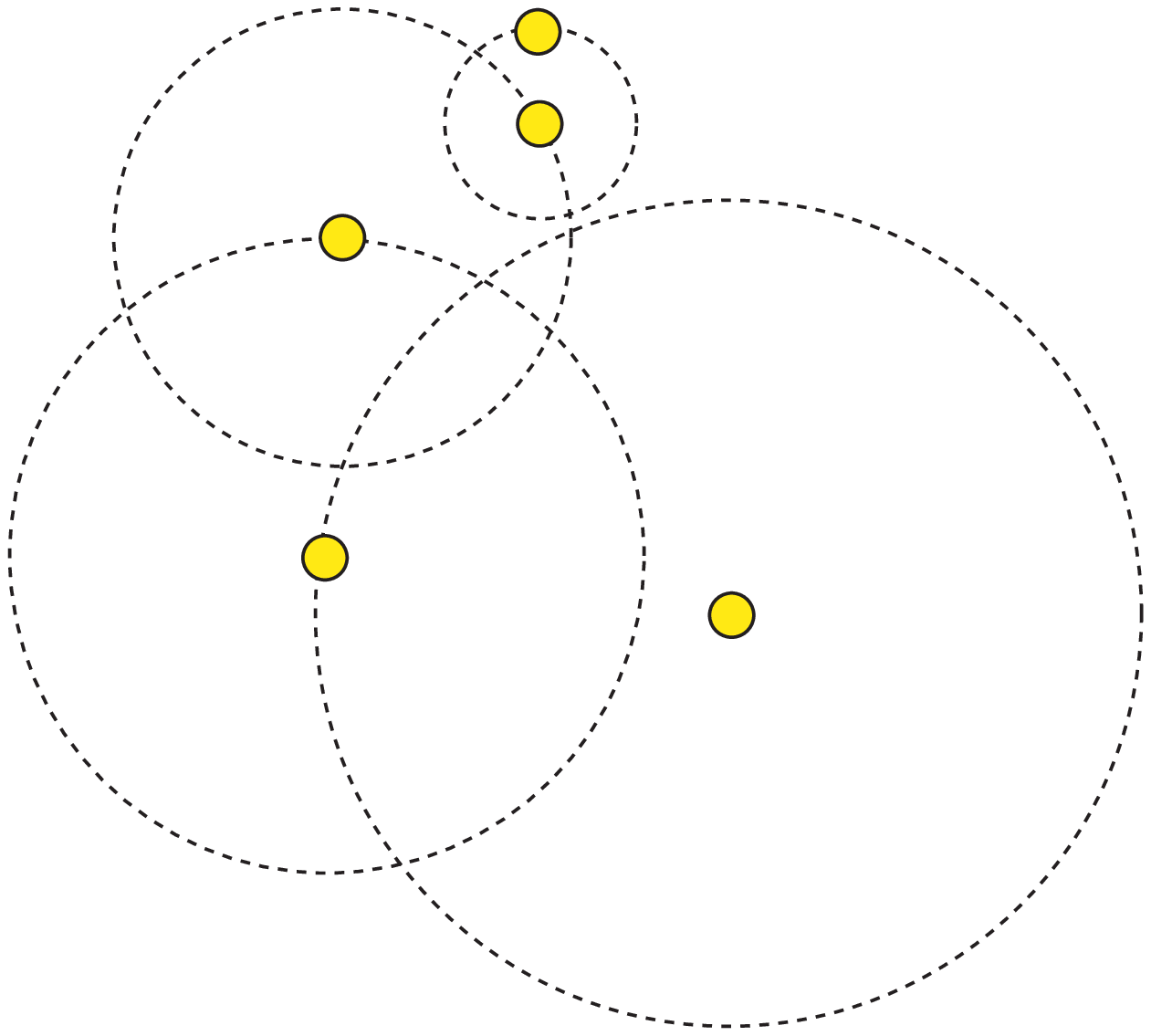,height=5cm}}
\caption{The Ptolemaean particle system corresponding to the $k=4$
Stern-Yi dyon chain. \label{epicycles}}}

In the corresponding particle mechanics picture we have one
particle for each node of the quiver, interacting with its two
nearest neighbors through (\ref{multiL12})-(\ref{multiL12}). The
equilibrium positions are given by equation (\ref{multieq}):
\begin{equation}
 \frac{\kappa_v}{2 |\sx_v - \sx_{v+1}|} -
 \frac{\kappa_{v-1}}{2|\sx_{v-1}-\sx_v|} = \theta_v \quad \mbox{for all
 } v=1,\ldots,k+1.
\end{equation}
where again $\kappa_0 \equiv 0$ and $\kappa_{k+1} \equiv 0$. This
is equivalent to
\begin{eqnarray*}
 |\sx_{k}-\sx_{k+1}| &=& - \frac{\kappa_k}{2 \theta_{k+1}} \\
 |\sx_{k-1}-\sx_{k}| &=& - \frac{\kappa_{k-1}}{2 (\theta_{k}+\theta_{k+1})} \\
 &\cdots& \\
 |\sx_{1}-\sx_{2}| &=& - \frac{\kappa_1}{2 (\theta_2+\theta_3+\cdots+\theta_{k+1})} = \frac{\kappa_1}{2 \theta_{1}}
\end{eqnarray*}
So the low energy motion of these particles is on some sort of
Ptolemaean chain of epicycles (or more accurately epispheres), as
shown in fig.\ \ref{epicycles}. Without going into the formal
details, it is intuitively clear that upon quantization of this
chain, one will again find a direct product structure, resulting
in the above generating function for the ground states. This
picture also explains where the huge degeneracy of these states
comes from.

\section{Conclusions and discussion} \label{sec:conclusions}

We considered two seemingly very different pictures of BPS bound
states, one as a set of particles at equilibrium separations from
each other, the other as a fusion of D-branes at a single point of
space, and we saw how the two are continuously related in the
context of quantum quiver mechanics, by changing the string
coupling constant $g_s$ (with the single D-brane picture
corresponding to $g_s \to 0$). We illustrated how this duality can
be used to solve some quite nontrivial ground state counting
problems in multi-particle ``electron-monopole'' quantum
mechanics, and to count BPS degeneracies of certain dyons in
supersymmetric Yang-Mills theories. Recent mathematical results on
the cohomology of quiver varieties allowed us to give a general
degeneracy formula for all such systems described by quivers
without closed loops.

Strictly speaking, the quiver models discussed in this paper are
only accurate for small phase differences (needed to cleanly
separate the low energy string modes from the massive ones).
However, even for bigger phase differences, the models can in many
case still be expected to capture many of the qualitative features
of the physics, especially since the most dramatic qualitative
changes happen at marginal stability loci, where phase differences
vanish rather than being bigger. In particular, one can in many
cases expect the ground state counting to remain valid. This does
not necessarily mean we can keep things under control for
\emph{arbitrary} phase differences, since by running around in
moduli space, monodromies can occur which invalidate the quiver
picture even qualitatively. For example in the case of dyons in
$\CN=2$ $SU(2)$ Yang-Mills theory, running around the marginal
stability line will change the dimension vector relevant for the
description of the dyon, so at some point the original description
must have broken down. On the other hand, the ground state
counting should give identical results, so there must be a
symmetry between these different representations. To get a
satisfactory unified description of such situations, the framework
of \cite{Dcat} is needed.

A problem left unanswered in this paper is whether the precise
match between the ground state degeneracies in the Higgs and the
Coulomb regimes extends from the examples considered here to the
general case. We conjectured that this is indeed true. To prove
this, it would be sufficient to show that the supersymmetric
ground states in the Coulomb regime are either all fermionic or
all bosonic, as is the case in the Higgs regime, but we did not do
this.

Another open question is what can be said about quivers with
closed loops (and hence possibly with superpotentials). Our
counting results were all for quivers without closed loops, in
part because the physics of those is more transparent, and in part
because much of the mathematics of quivers with closed loops is
still unknown. It would be very interesting to extend our results
to those cases, especially since this includes the quivers
describing black hole states.

\acknowledgments

I would like to thank Ben Craps, Mike Douglas, Simeon Hellerman,
Kentaro Hori, Rob Myers, Greg Moore, Cumrun Vafa and Piljin Yi for
helpful conversations, and the Aspen Center for Physics, Harvard
University, the NHETC at Rutgers University, the Perimeter
Institute, and the Korean Institute for Advanced Study for
stimulating hospitality.

%%%%%%%%%%%%%%%%%%%%%%%%%%%%%%%%%%%%%%%%%%%%%%%%%%%%%%%%%%%%%%%%%%
%%%%%%%%%%%%%%%%%%%%%%%%%%%%%%%%%%%%%%%%%%%%%%%%%%%%%%%%%%%%%%%%%%

\appendix
\setcounter{equation}{0}

\section{Notations and conventions} \label{conventions}

Our metric signature is $(-+++)$. Spinors with indices down
transform in the ${\bf 2}$ of the spatial $SO(3)$, spinors with
indices up in the ${\bf \bar{2}}$. The unbarred spinors appearing
in this paper all have indices down, the barred ones indices up.
Barred and unbarred spinors are related through complex
conjugation: $(\psi_\alpha)^* \equiv \bar{\psi}^\alpha$. We use
the following notations:
\begin{eqnarray}
 && \bar{\psi} \chi = \bar{\psi}^\alpha \chi_\alpha = - \chi_\alpha
 \bar{\psi}^\alpha = - \chi \bar{\psi} \\
 &&\bar{\psi} \sigma^i \chi = \bar{\psi}^\alpha
 {{\sigma^i}_\alpha}^\beta \chi_\beta \\
 && \epsilon^{\alpha \beta} = - \epsilon^{\beta \alpha}, \quad
 \epsilon_{\alpha \gamma} \epsilon^{\gamma \beta} =
 {\delta_\alpha}^\beta , \quad \epsilon^{12} = 1, \quad \epsilon_{1 2} =
 -1 \\
 && (\epsilon \psi)^\alpha = \epsilon^{\alpha\beta} \psi_\beta \label{epsup} \\
 && (\bar{\psi} \epsilon)_\alpha = \bar{\psi}^\beta \epsilon_{\beta
 \alpha} \label{epsdown}
\end{eqnarray}
We do not define index lowering or raising; instead we always
write the appropriate $\epsilon$ explicitly, as in
(\ref{epsup})-(\ref{epsdown}). So for instance an invariant
contraction of two lower index spinors will look like $\psi
\epsilon \chi = \psi_\alpha \epsilon^{\alpha \beta} \chi_\beta$.
The $\sigma^i$ are the Pauli matrices:
\begin{equation}
 \sigma^1 =
 \left(
 \begin{array}{cc}
  0 & 1 \\
  1 & 0
 \end{array}
 \right), \quad
 \sigma^2 =
 \left(
 \begin{array}{cc}
  0 & -i \\
  i & 0
 \end{array}
 \right), \quad
 \sigma^3 =
 \left(
 \begin{array}{cc}
  1 & 0 \\
  0 & -1
 \end{array}
 \right).
\end{equation}

\section{Supersymmetry transformations for the $U(1)$ case} \label{app:susy}

The relevant supersymmetry transformations for the different forms
of the Lagrangian describing the example of section
\ref{sec:example} are closely related. We give those for the
relative Lagrangian (\ref{relLV})-(\ref{relLC}), which we copy
here for convenience:
\begin{eqnarray}
 L_{rel} &=& \frac{\mu}{2} \left( {\dot{\sx}}^2 + {D}^2
 + 2 i \bar{\lambda} \dot{\lambda} \right) - \theta D  \\
 && +  |\CD_t \phi^a|^2 - \left( \sx^2 + D \right) |\phi^a|^2 +
 |F^a|^2+ i \, \bar{\psi^a} \CD_t \psi^a \nonumber \\
 &&
 - \bar{\psi^a} \, \sx \cdot \ssigma \, \psi^a
 - i \sqrt{2} (\bar{\phi}^a \psi^a \epsilon \lambda -
 \bar{\lambda} \epsilon \bar{\psi}^a \phi^a) \, ,
\end{eqnarray}
The corresponding supersymmetry transformations are:
\begin{eqnarray}
 \delta A &=& i \, \bar{\lambda} \, \xi - i
 \, \bar{\xi} \,  \lambda \label{rule1} \\
 \delta \sx &=& i \, \bar{\lambda} \, \ssigma \, \xi - i
 \, \bar{\xi} \, \ssigma \, \lambda \\
 \delta \lambda &=& \dot{\sx} \cdot \ssigma \, \xi
 + i \, D \, \xi  \\
 \delta D &=& - \dot{\bar{\lambda}} \, \xi
  - \bar{\xi} \, \dot{\lambda} \\
 \delta \phi^a &=& \sqrt{2} \, \epsilon \xi \, \psi^a  \\
 \delta \psi^a &=& - i \sqrt{2} \, \bar{\xi} \epsilon \, \CD_t \phi^a
 - \sqrt{2} \, \sx \cdot \ssigma \, \bar{\xi} \epsilon \, \phi^a
 + \sqrt{2} \, \xi \, F^a \\
 \delta F^a &=& - i \sqrt{2} \, \bar{\xi} \, \CD_t \psi^a
 + \sqrt{2} \, \bar{\xi} \, \ssigma \, \psi^a \cdot \sx
 - 2 i \, \bar{\xi} \epsilon \, \bar{\lambda} \, \phi^a
 \label{finalrule}
\end{eqnarray}

\section{General quiver mechanics Lagrangian}
\label{app:genquiv}

A quiver $Q$ with nodes $v \in V$, arrows $a \in A$, and dimension
vector $N=(N_v)_{v \in V}$ corresponds to an $\CN=1$, $d=4$ gauge
theory, or, in our setting, to an $\CN=4$, $d=1$ matrix model,
obtained by dimensional reduction from $d=4$. To each node $v$, we
associate a linear (a.k.a.\ vector) multiplet
$(A_v,X_v^i,\lambda_v,D_v)$, $i=1,2,3$, $v \in V$, with gauge
group $U(N_v)$, and to each arrow $a:v \to w$, we associate a
bifundamental chiral multiplet $(\phi^a,\psi^a,F^a)$, transforming
in the $(\boldsymbol{\bar{N}_v},\boldsymbol{N_w})$ of $U(N_v)
\times U(N_w)$. The corresponding $\CN=4$, $d=1$ Lagrangian is, in
units with $l_s = 2 \pi \alpha' = 1$:
\begin{equation}
 L = L_V + L_{FI} + L_C + L_I + L_W
\end{equation}
with
\begin{eqnarray*}
 L_V &=& \sum_v \frac{m_v}{2} \tr \biggl( (D_t X^i_v)^2 + {D_v}^2 - \frac{1}{2}
 [X^i,X^j]^2
 + \frac{}{} 2 i {\lambda_v}^\dagger \CD_t \lambda_v
 - 2 {\lambda_v}^\dagger \sigma^i [X^i,\lambda_v] \biggr), \\
 L_{FI} &=& \sum_v - \theta_v \tr D_v \\
 L_C &=& \sum_a \tr \biggl( |\CD_t \phi^a|^2
 + |F^a|^2 + i \, {\psi^a}^\dagger \CD_t \psi^a \biggr) \\
 L_I &=& \sum_{a:v \to w} - \tr \biggl(
 |X_w^i \phi^a - \phi^a X_v^i|^2
 + {\phi^a}^\dagger (D_w \phi^a - \phi^a D_v)
 + {\psi^a}^\dagger \sigma^i (X_w^i \psi^a - \psi^a X_v^i) \\
 && - i \sqrt{2} \left( ({\phi^a}^\dagger \lambda_w - \lambda_v
 {\phi^a}^\dagger) \epsilon \psi^a - {\psi^a}^\dagger \epsilon ({\lambda_w}^\dagger \phi^a -
 \phi^a {\lambda_v}^\dagger ) \right) \biggr) \\
 L_W &=& \sum_a \tr \biggl(\frac{\partial W}{\partial \phi^a} \, F^a
 + \mbox{h.c.} \biggr) + \frac{1}{2} \sum_{a,b} \tr \biggl(
 \frac{\partial^2 W}{\partial \phi^a \partial \phi^b} \, \psi^a
 \epsilon \psi^b + \mbox{h.c.} \biggr)
\end{eqnarray*}
where, for $a:v \to w$:
\begin{eqnarray*}
 \CD_t \phi^a &=& \partial_t \phi^a + i \, A_w \phi^a - i \phi^a A_v \\
 \CD_t X_v^i &=& \partial_t X_v^i + i [A_v,X^i_v]
\end{eqnarray*}
and similarly for the superpartners. The parameters $\theta_v$ in
$L_{FI}$ are the Fayet-Iliopoulos parameters, which for the
D-brane model are given by (\ref{FIgen}). A (gauge invariant)
holomorphic superpotential $W(\phi)$ can only appear if the quiver
has no closed loops.

The supersymmetry transformations are
\begin{eqnarray*}
 \delta A_v &=& i \, \bar{\lambda}_v \, \xi - i
 \, \bar{\xi} \,  \lambda_v \\
 \delta X^i_v &=& i \, \bar{\lambda}_v \, \sigma^i \, \xi - i
 \, \bar{\xi} \, \sigma^i \, \lambda_v \\
 \delta \lambda_v &=&  \CD_t X^i_v \sigma^i \, \xi
 + \frac{1}{2} \epsilon^{ijk} [X_v^i,X_v^j] \sigma^k \xi
 + i \, D_v \, \xi  \\
 \delta D &=& - \CD_t \bar{\lambda}_v \, \xi - i \,
 [X^i_v,\bar{\lambda}_v] \sigma^i \xi - \bar{\xi} \, \CD_t \lambda_v
 - i \, \bar{\xi} \sigma^i [X^i_v,\lambda_v] \\
  \delta \phi^a &=& \sqrt{2} \epsilon \xi \psi^a \\
 \delta \psi^a &=& - i \sqrt{2} \, \bar{\xi} \epsilon \, \CD_t \phi^a
 - \sqrt{2} \, \sigma^i \, \bar{\xi} \epsilon (X^i_w \phi^a - \phi^a
 X^i_v) + \sqrt{2} \, \xi \, F^a \\
 \delta F^a &=& - i \sqrt{2} \, \bar{\xi} \, \CD_t \psi^a
 + \sqrt{2} \, \bar{\xi} \, \sigma^i (X^i_w \psi^a - \psi^a X^i_v)
 - 2 i \, \bar{\xi}\epsilon  (\bar{\lambda}_w \phi^a - \phi^a \bar{\lambda}_v)
\end{eqnarray*}

When the vector multiplets are restricted to diagonal matrices,
the Lagrangian components simplify to
\begin{eqnarray*}
 L_V &=& \sum_v \sum_{n=1}^{N_v}\frac{m_v}{2} \left( (\dot{x}^i_{v,n})^2 + (D_{v,n})^2
 + 2 i {\bar{\lambda}_{v,n}} \dot{\lambda}_{v,n} \right) \, , \\
 L_{FI} &=& - \sum_v  \sum_{n=1}^{N_v} \theta_v  D_{v,n} \\
 L_C &=& \sum_a \tr \bigl( |\CD_t \phi^a|^2
 + |F^a|^2 + i \, {\psi^a}^\dagger \CD_t \psi^a \bigr) \\
 L_I &=& - \sum_{a:v \to w} \sum_{n,m=1}^{N_v,N_w}
 \biggl( \left( \frac{}{} (x^i_{w,m}-x^i_{v,n})^2 + D_{w,m}-D_{v,n} \right)
 |\phi^a_{mn}|^2\\
  && + (x^i_{w,m}-x^i_{v,n}) \overline{\psi^a_{mn}} \sigma^i \psi^a_{mn}\\
 && - i \sqrt{2} \left( \frac{}{} \, \overline{\phi^a_{mn}}  (\lambda_{w,m} -
 \lambda_{v,n})\epsilon \psi^a_{mn} - \overline{\psi^a_{mn}} \epsilon (\bar{\lambda}_{w,m} -
 \bar{\lambda}_{v,n})  \phi^a_{mn} \right) \biggr) \\
 L_W &=& \sum_a \tr \biggl(\frac{\partial W}{\partial \phi^a} \, F^a
 + \mbox{h.c.} \biggr) + \frac{1}{2} \sum_{a,b} \tr \biggl(
 \frac{\partial^2 W}{\partial \phi^a \partial \phi^b} \, \psi^a
 \epsilon \psi^b + \mbox{h.c.} \biggr)
\end{eqnarray*}

%%%%%%%%%%%%%%%%%%%%%%%%%%%%%%%%%%%%%%%%%%%%%%%%%%%%%%%%%%%%%%%%%%%%%%%%
%%%%%%%%%%%%%%%%%%%%%%%%%%%%%%%%%%%%%%%%%%%%%%%%%%%%%%%%%%%%%%%%%%%%%%%%

\newcommand{\dgga}[1]{\href{http://xxx.lanl.gov/abs/dg-ga/#1}{\tt
dg-ga/#1}}
\renewcommand\baselinestretch{1.08}\normalsize

\newcommand{\mathdg}[1]{\href{http://xxx.lanl.gov/abs/math.DG/#1}{\tt
math.DG/#1}}
\renewcommand\baselinestretch{1.08}

\end{document}